\documentclass[a4paper,fleqn]{cas-dc}

\usepackage{relsize}
\usepackage{titlesec} 
\usepackage{graphicx}

\titleformat*{\section}{\small\bfseries}
\titleformat*{\subsection}{\small\bfseries}
\titleformat*{\subsubsection}{\small\bfseries}

\usepackage[numbers,sort&compress,longnamesfirst]{natbib}
\usepackage{xcolor}  
\usepackage{soul}
\definecolor{authorcolor}{rgb}{0,0,0}  
\definecolor{myblue}{RGB}{0, 0, 0}
\definecolor{myblue2}{RGB}{188, 231, 245}
\definecolor{myblue1}{RGB}{255, 255, 255}
\usepackage{url}
\usepackage{amsmath,amsfonts}
\usepackage{hyperref}
\hypersetup{
    colorlinks=true, 
    linkcolor=myblue,  
    citecolor=myblue,  
    urlcolor=myblue  
}
\usepackage{newtxtext} 
\usepackage{caption}
\usepackage{float} %
\usepackage{lipsum} %
\usepackage{tcolorbox} %
\newtcolorbox{mybox}{
    colback=white, %
    colframe=black, %
    boxrule=1pt, %
    arc=0pt, %
    left=8pt, %
    right=8pt, 
    top=6pt, 
    bottom=6pt, 
    fontupper=\rmfamily\footnotesize 
}

\usepackage[ruled]{algorithm2e}

\usepackage{fancyhdr}
\def\tsc#1{\csdef{#1}{\textsc{\lowercase{#1}}\xspace}}
\tsc{WGM}
\tsc{QE}
\tsc{EP}
\tsc{PMS}
\tsc{BEC}
\tsc{DE}

\begin{document}
\let\WriteBookmarks\relax
\def\floatpagepagefraction{1}
\def\textpagefraction{.001}


\title[mode = title]{\textcolor{myblue}{A data-driven approach for topology correction in low voltage distribution networks with PVs}}



%
\author[1]{\textcolor{authorcolor}{Dong Liu}}
\credit{Conceptualisation, Methodology, Software, Validation, Writing - original draft}

\author[2]{\textcolor{authorcolor}{Sander Timmerman}}
\credit{Software, Validation}

\author[2,3]{\textcolor{authorcolor}{Yu Xiang}}
\credit{Writing - review \& editing}

\author[2]{\textcolor{authorcolor}{Ensieh Hosseini}}
\credit{Software}

\author[1]{\textcolor{authorcolor}{Peter Palensky}}
\credit{Funding acquisition}

\author[1]{\textcolor{authorcolor}{Pedro P. Vergara}}[orcid=0000-0003-0852-0169]
\cormark[1]
\ead{P.P.VergaraBarrios@tudelft.nl}
\credit{Writing - review \& editing, Supervision, Funding acquisition}

\affiliation[1]{organization={Intelligent Electrical Power Grids (IEPG) Group},
    addressline={Delft University of Technology}, 
    postcode={2628CD}, 
    country={The Netherlands}}
 
\affiliation[2]{organization={Alliander N.V.},
    postcode={6812 AH}, 
    country={The Netherlands}}
    
\affiliation[3]{organization={Electrical Energy Systems group},
    addressline={Eindhoven University of Technology}, 
    postcode={5612AZ}, 
    country={The Netherlands}}
\cortext[cor1]{Corresponding author}



\begin{abstract}
\textcolor{myblue}{To correct the outdated and incomplete topology of low-voltage distribution networks (LVDNs) using solely voltage magnitudes, a data-driven approach is developed by integrating machine learning algorithms with correlation analysis. Unlike conventional offline approaches, the proposed method leverages up-to-date voltage magnitude measurements within a unified framework, enabling the same dataset to be consistently utilized across all processing stages without repeated data preprocessing. Specifically, switch states are identified via supervised learning, while user–feeder connections and customer phase labels are refined using a modified hierarchical clustering algorithm. To address the similarity among smart meter (SM) data induced by distributed photovoltaic (PV) systems, a time-based SM data selection strategy is incorporated into the correlation analysis. The feasibility and robustness of the proposed approach are validated using modified real-world LVDNs and multiple incomplete SM datasets collected from customers in the Netherlands. The results demonstrate that the time-based SM data selection strategy effectively mitigates the impact of PV-induced similarity on phase identification, and that the corrected topology not only improves network observability but also supports distribution system operators in load balancing and PV integration.}
\vspace{-\baselineskip}
\end{abstract}


\maketitle

\section{Introduction}
\subsection{\textcolor{myblue}{Motivation}}
\textcolor{myblue}{Most existing phase balancing and topology reconfiguration problems are formulated as mixed-integer optimization problems that depend on network topologies~\cite{10098964,11017695,10571996}. However, these topologies are often inaccurate and outdated for distribution system operators~(DSOs) due to missing recordings, topology maintenance and reconfiguration, such as congestion management ~\cite{vanin2024phase}. Thus, the topology of the low-voltage distribution network (LVDN) needs to be checked and corrected when it is outdated. The increasing uncertainty of distributed energy resources (DERs), including household photovoltaic (PV), heating pumps, etc., impacts the frequency of topology reconfiguration and challenges the correction of the low-voltage distribution network topology~\cite{10026490, 10347462, 10475702}. Moreover, the available smart meter (SM) datasets are often limited due to privacy concerns and random communication channel failure, challenging the topology correction~\cite{9696306, costa2022identification, dande2025consumer}. Synthetic European networks and benchmark models presented in~\cite{birchfield2016grid,2020Non} are benchmarks for research but insufficient to represent the diversity of European LVDNs for practical use by DSOs (e.g., state estimation). Thus, practical topology identification and correction approaches are required for real-time topology updating for active management of LVDNs.}

\subsection{\textcolor{myblue}{Literature review}}
\begin{figure*}
\centering
\includegraphics[width=0.99\textwidth,height=0.3\textwidth]{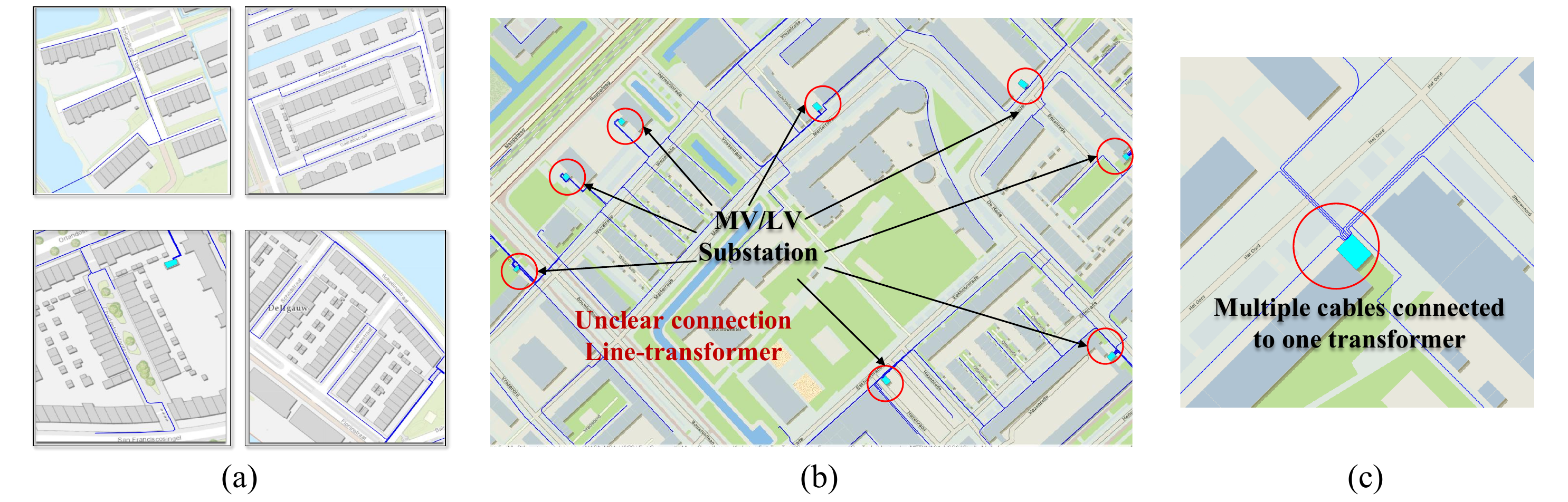}
\caption{\textnormal{\textcolor{myblue}{Deployment patterns of LVNDs: (a) feeder deployment under one street, (b) distribution of MV/LV transformers and (c) underground feeders connected to the same transformer.}}}
\label{Fig1}
\vspace{0.3cm}
\end{figure*}

\textcolor{myblue}{The topology identification problem in distribution networks~(DNs) could be briefly classified into three categories: (1) switch state identification, (2) connection lines and their parameters identification, and (3) phase identification. The switch state identification was taken as a classification problem in~\cite{9758816}, solved by a trained deep neural network and providing quick estimations under the selected features. To recover accurate connection lines between transformers and customers, a probabilistic graphical model-based method was developed in~\cite{8456535} by restricting the propagation of SM data error. In~\cite{liu2023hybrid}, a hybrid data-driven approach integrating a partial correlation analysis strategy and a linear regression model is proposed to estimate topology from limited SM data. Based on an alternating direction method of multipliers, a robust topology identification method is proposed to estimate the line connection and their parameters using $\mu$-PMUs and SM datasets~\cite{9729419}. A two-step approach is introduced to identify the topology of LVDNs with multiple meter types, which provide different resolution datasets, mitigating the dependence on the SMs~\cite{10400804}. Related to phase identification, current approaches rely on voltage magnitude data and demand profiles~\cite{garcia2025data}. Since the correlation of voltage magnitude data from the same phase is stronger than that from different phases, correlation analysis is normally used. Given voltage magnitude datasets, spectral clustering and $k$-means clustering approaches are adopted to identify customer phase label~\cite{8955915}. The mixed integer linear programming model in~\cite{9409966} is constructed to identify the phase labels of single-phase customers by minimizing the gap between transformers' measurements and the estimated data (e.g., power and voltage magnitude) in each phase. Saliency analysis (SA) and statistical tests are combined in \cite{jimenez2020phase} to guarantee the identification accuracy of the data pairs with weak correlations. \textcolor{myblue}{Fig. \ref{Fig1}} presents several underground feeder deployment patterns in the Netherlands~\cite{web1}, which indicates the challenges of LVDN topology identification and correction in practice, e.g., multiple cables under the same street. In addition, some of the above approaches are sensitive to missing data and measurement errors, e.g., SA-based approaches. Furthermore, the user-feeder connection identification issue is rarely considered in the aforementioned papers due to the assumption that one main feeder connects to the transformer.}


\begin{table*}[!t]
\begin{mybox}
\textit{\textbf{Acronyms}}\\
\begin{tabular}{@{}ll@{}}

$\textbf{MV/LV}$ & Medium/Low voltage\\
$\textbf{LVDN}$ & Low-voltage distribution network\\
$\textbf{DNs}$ & Distribution networks\\
$\textbf{SM}$ & Smart meter\\
$\textbf{DSO}$ & Distribution system operator\\
$\textbf{DERs}$ & Distribution energy resources\\
$\textbf{SA}$ & Saliency analysis\\
$\textbf{PV}$ & Photovoltaic\\
$\textbf{EV}$ & Electical vehicles\\
$\textbf{HC}$ & Hierarchical clustering algorithm\\
$\textbf{RF}$ & Random forest algorithm\\
$\textbf{KNN}$ & K-nearest neighbors algorithm\\
$\textbf{GNB}$ & Gaussian Naive Bayes algorithm\\
$\textbf{SVM}$ & Support vector machine algorithm\\
$\textbf{ANN}$ & Artificial neural network\\
$\textbf{PCC}$ & Pearson correlation coefficient\\

\end{tabular}
\hfill
\begin{tabular}{@{}ll@{}}
\hspace{1.2cm} $\textbf{MFP}$ & Modified PCC based Fisher z-transformation\\
\hspace{1.2cm} $\textbf{GIS}$ & Geographic Information System\\
\hspace{1.2cm} \textbf{CN1} & Network with two complex-structured feeders\\
\hspace{1.2cm} \textbf{CN2} & Network with one simple- and one \\
\hspace{5cm}& complex-structured feeder\\
\hspace{1.2cm} \textbf{SN} & Network with two simple-structured feeders\\
\hspace{1.2cm} \textbf{SNL} & Network with two longer simple-structured feeders\\
\hspace{1.2cm} \textbf{SNB} & Network with two simple-structured feeders \\
\hspace{5cm}& with more branches\\
\hspace{1.2cm} \textbf{CC\_PV} & CN1 with PV at each house\\
\hspace{1.2cm} \textbf{SC\_PV} & SN with PV at each house\\
\hspace{1.2cm} \textbf{CC} & Complex feeder from CN1\\
\hspace{1.2cm} \textbf{SC} & Simple feeder from SN\\
\hspace{1.2cm} \textbf{CC\_3\_NU} & Complex feeder with three-phase non-uniform loads\\
\hspace{1.2cm} \textbf{SC\_3\_NU} & Simple feeder with three-phase non-uniform loads\\
\end{tabular}

\end{mybox}
\end{table*}


\textcolor{myblue}{To address this issue, based on a modified genetic algorithm, a novel data-driven phase identification approach is introduced to identify the phase labels under incomplete datasets~\cite{jimenez2021new}. Based on incomplete and asynchronous SM data, an optimization-based data-driven approach is constructed to recover the approximate topologies in LVDNs with open datasets~\cite{liu2025topology}. Voltage magnitude fluctuations among customers connected to different phases may exhibit similarities due to variations induced by the medium-voltage~(MV) network or similar load patterns. To mitigate this effect, a high-pass filter is applied during time-series data pre-processing in the phase identification approach \cite{hosseini2020machine}. Additionally, to mitigate the impact of electrical distance on the clustering of SM time-series data, a power-band-based strategy is proposed to select data sets, thereby improving the precision of phase identification and transformer-to-meter matching~\cite{10533846}. An optimisation model-based approach was introduced in \cite{10742905} to estimate customer phase labels and switch states in LVDNs with limited available recordings. SA strategies were integrated to extract underlying features, thereby enhancing the accuracy of phase-label identification~\cite{7604110}. Nevertheless, the impacts of household DERs have not been sufficiently addressed in the above research. With the increasing penetration of household PV installations, the similarity among household voltage magnitudes is expected to increase significantly. Specifically, within the same LVDN, PV systems are exposed to similar solar irradiance conditions, resulting in highly correlated PV power generation patterns. This leads to similar voltage magnitude distributions among households, challenging the feasibility of correlation analysis-based approaches.}

\textcolor{myblue}{Topology identification generally aims to determine the structure of an LVDN when the network topology is completely unknown. In this setting, the potentially valuable information contained in previously recorded topologies is typically not utilised. In contrast, topology correction focuses on updating only the incorrect or outdated components of an existing topology by leveraging limited recent data while fully exploiting the information available in historical topology records. While topology identification has attracted significant research attention in recent years, topology correction remains relatively underexplored. To dynamically update the connection and the parameters after topology configuration, a graph fusion network-based approach presented in \cite{10734663} is used to extract correct connection lines. However, the adjacency matrix is also required. For networks with low observability, a distance matrix is introduced in~\cite{10612997} to correct the topology and is also effective in detecting latent nodes positioned between observed nodes along network paths (i.e., intermediate latent nodes rather than endpoints). A physics-informed neural network is constructed to estimate the topology by representing the nodes without meters, whose parameters are adjusted by a reinforcement learning-based approach~\cite{10138375}. A data-driven approach was introduced in \cite{7098424} to correct the connection between meters and transformers in GIS systems using voltage datasets. A similar work in~\cite{wang2023data} established a data-driven model based on the impedance matrix to correct topology and identify the impedance of connection lines using SM data. A data-driven approach in \cite{11063399} is to infer latent nodes and line parameters for outdated topologies using power and voltage magnitude profiles. Nevertheless, different approaches often rely on different assumptions and require different types of input SM data. As a result, DSOs need to prepare multiple datasets to perform separate identification tasks.}

\subsection{\textcolor{myblue}{Contributions}}
To overcome these limitations, we propose a data-driven approach for topology correction in LVDNs using incomplete voltage data. Table \ref{table1} highlights the key differences between the proposed approach and existing approaches. By relying solely on voltage magnitude data, this approach preserves privacy and minimizes SM updates, as it releases the need for phase angle measurements from households. The main contributions are summarized as follows:

\begin{itemize}

\begin{table*}
\centering
\caption{\textcolor{myblue}{Comparison of representative topology identification and correction research in LVDNs.}}
\label{table1}
\resizebox{0.99\textwidth}{!}{  
\begin{tabular}{cc|cccc|ccc|cccc}
\hline
\hline
\multirow{3}{*}{\textbf{Method}} &
\multirow{3}{*}{\textbf{Ref.}} &
\multicolumn{4}{|c}{\textbf{Input Data}} & 
\multicolumn{3}{|c}{\textbf{Robustness}} & 
\multicolumn{4}{|c}{\textbf{Output}} \\ 
& & \multirow{2}{*}{$V$} & \multirow{2}{*}{$P/I$} & Limited & GIS & DERs & SM & Missed & Switch & user-feeder & Phase & \textcolor{myblue}{latent}\\ 
&  & & & topology & data & impact & errors &  data&  state& connections & labels & \textcolor{myblue}{nodes}\\ 
\hline
\multirow{3}{*}{ANN}  & \cite{9758816} & $\times$ & $\checkmark$ & $\times$ & $\times$ & $\checkmark$ & $\times$ &  $\times$ &  $\checkmark$ & $\times$ & $\times$ & $\times$\\
  & \cite{10138375} & $\checkmark$ & $\checkmark$ &$\checkmark$  & $\times$ & $\times$ & $\times$  & $\checkmark$ & $\times$ & $\checkmark$ & $\times$ & $\times$\\
& \cite{10734663} & $\checkmark$ & $\checkmark$ &$\checkmark$  & $\times$ & $\times$ & $\checkmark$  & $\times$ & $\times$ & $\checkmark$ & $\times$ & $\checkmark$\\
  
\hline
\multirow{2}{*}{Linear regression}  & \cite{8456535} & $\checkmark$ & $\times$ & $\checkmark$ & $\times$ & $\checkmark$ & $\checkmark$ & $\times$ & $\times$ & $\checkmark$ & $\times$ & $\times$\\
  & \cite{cunha2020automated} &  $\checkmark$ & $\checkmark$ & $\checkmark$ & $\checkmark$ & $\checkmark$& $\checkmark$ & $\checkmark$ & $\times$ &$\checkmark$ & $\checkmark$ & $\times$\\
\hline
\multirow{3}{*}{Optimization based}  & \cite{liu2025topology} & $\checkmark$ & $\checkmark$ & $\times$ & $\checkmark$ & $\times$ & $\checkmark$ & $\checkmark$ & $\times$ & $\checkmark$ & $\times$ & $\times$\\
& \cite{9409966} & $\checkmark$ & $\checkmark$ & $\checkmark$ & $\times$ & $\checkmark$ & $\checkmark$ & $\times$ & $\times$ & $\times$ & $\checkmark$ & $\times$\\
& \cite{11063399} & $\checkmark$ & $\checkmark$ & $\checkmark$ & $\times$ & $\checkmark$ & $\checkmark$ & $\times$ & $\times$ & $\checkmark$ & $\checkmark$ & $\times$\\

\hline
\multirow{4}{*}{clustering-based}  & \cite{8955915} &$\checkmark$ &$\times$ & $\times$ & $\times$ & $\times$ & $\times$ & $\times$ & $\times$& $\times$ & $\checkmark$ & $\times$\\
& \cite{10533846} & $\checkmark$ & $\checkmark$ & $\times$ & $\times$ & $\times$&$\times$ & $\times$ & $\times$& $\checkmark$ & $\checkmark$ & $\times$\\
& \cite{7098424} & $\checkmark$ & $\checkmark$ & $\times$ & $\checkmark$ & $\times$&$\times$ & $\times$ & $\times$& $\checkmark$ & $\checkmark$ & $\times$\\
& \cite{wang2023data} & $\checkmark$ & $\checkmark$ & $\checkmark$ & $\times$ & $\times$& $\checkmark$ & $\times$ & $\times$ & $\checkmark$ & $\times$ & $\times$\\

\hline
\multirow{2}{*}{Heuristic Algorithms} & \cite{jimenez2021new} & $\times$ & $\checkmark$ &$\times$ & $\times$ & $\checkmark$  & $\checkmark$  & $\checkmark$  & $\times$ & $\times$ & $\checkmark$ & $\times$\\
& \cite{10612997} & $\checkmark$ & $\checkmark$ & $\checkmark$ & $\times$ & $\times$& $\checkmark$ & $\times$ & $\times$ & $\checkmark$ & $\times$ & $\checkmark$\\

\hline
Our Method &  & $\checkmark$ & $\times$ & $\checkmark$  & $\times$ &  $\checkmark$ & $\checkmark$ & $\checkmark$  & $\checkmark$  & $\checkmark$ & $\checkmark$ & $\times$\\
\hline
\hline
\end{tabular}
}
\\
\footnotesize{\textcolor{myblue}{Note: Topology reconfiguration studies are not included in this comparison, as they assume a known network topology and focus on optimal network operation (e.g., loss reduction and voltage regulation), rather than topology identification or correction.}}
\end{table*}

\item \textcolor{myblue}{A systematic data-driven topology correction framework is proposed to update outdated LVDN topology records through three sequential steps: (1) switch state identification, (2) user–feeder connection correction, and (3) phase label identification. The proposed framework relies solely on voltage magnitude measurements from SMs, enabling the identification and correction of multiple topology components using a unified input dataset. Compared with conventional approaches that require complete SM datasets or different types of measurements for separate tasks, the proposed method allows DSOs to correct topology information using limited and noisy SM voltage data, even when household measurements are incomplete.}

\begin{figure*}
\centering
\includegraphics[width=0.99\textwidth,height=0.31\textwidth]{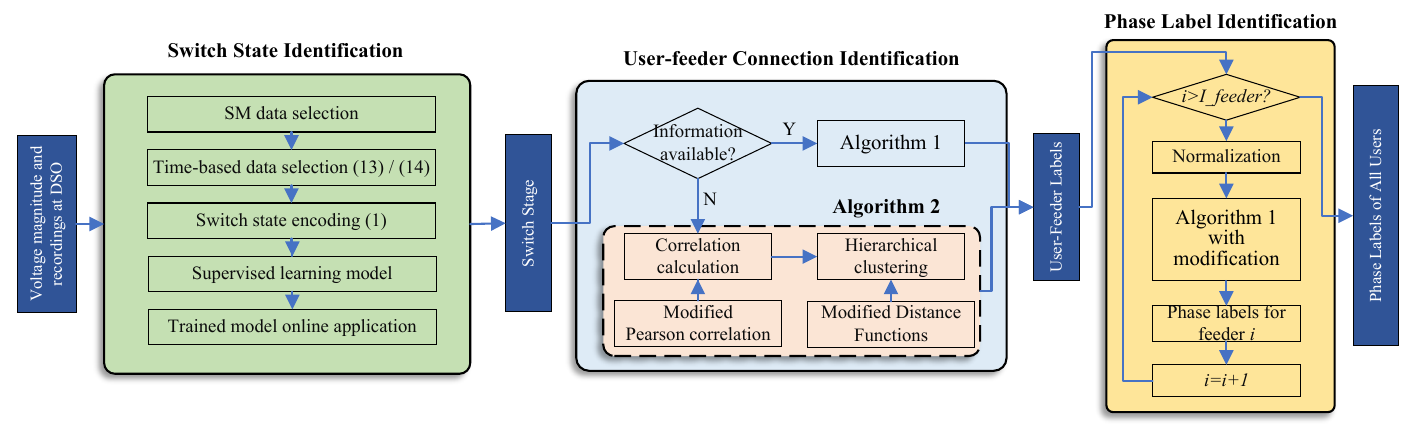}
\caption{\textnormal{\textcolor{myblue}{Proposed topology correction approach composed of three steps: (1) switch state identification, (2) user-feeder connection identification and (3) phase labels identification.}}}
\vspace{-0.3cm}
\label{Fig.2}
\end{figure*}

\item \textcolor{myblue}{A switch state identification strategy is developed by encoding the operational states of switch bars (including multiple switchgears) using label encoding rules. In the topology correction context, historical topology records and network operation logs provide labelled samples corresponding to different switch configurations. This formulation converts the identification of multiple switchgear states into a supervised classification problem that can be efficiently solved using existing machine learning algorithms. Furthermore, a modified distance function is incorporated into traditional hierarchical clustering (HC)–based correlation analysis to enhance the distinction between geographically close loads belonging to different phases or cables, thereby improving the accuracy of user–feeder and phase identification.}

\item \textcolor{myblue}{To mitigate the influence of PV power on voltage-based correlation analysis, a time-based SM data selection strategy is proposed to extract voltage measurements during periods when household loads dominate voltage variations, and PV generation remains lower than the local demand. If PV generation forecasts are available, the selection can be refined by identifying intervals with low PV injection. Otherwise, voltage data after sunset are used as a practical utilization. This strategy reduces the distortion of correlation patterns caused by PV generation and improves the accuracy of topology correction without requiring explicit PV modeling.}
\end{itemize}

The remainder of this paper is organised as follows: Section~\textcolor{myblue}{\ref{section2}} details the framework of the proposed approach, including identification of switch states, user-feeder connections, phase labels, and voltage data pre-processing. Section~\textcolor{myblue}{\ref{case}} presents the case studies and results. Finally, Section~\textcolor{myblue}{\ref{Conclusion}} summarizes the conclusions of this work.

\section{\textcolor{myblue}{Topology correction approach}}
\label{section2}
The proposed topology correction approach is illustrated in  Fig.~\textcolor{myblue}{\ref{Fig.2}}, which is composed of three steps: (1) switch state identification, (2) user-feeder connection identification and (3) phase label identification. \textcolor{myblue}{An illustrative example is presented in Fig.~\ref{Fig.5}, which shows the outputs of each step. The process is triggered when the latest voltage magnitude measurements become available or an outdated topology is detected.} The input includes customer voltage magnitude and incomplete recordings of the user-feeder connection, and the output is the corrected topology with accurate customer's user feeder connections and phase labels, including single-phase and three-phase customers.

\begin{figure}
\centering
\includegraphics[width=0.49\textwidth,height=0.205\textwidth]{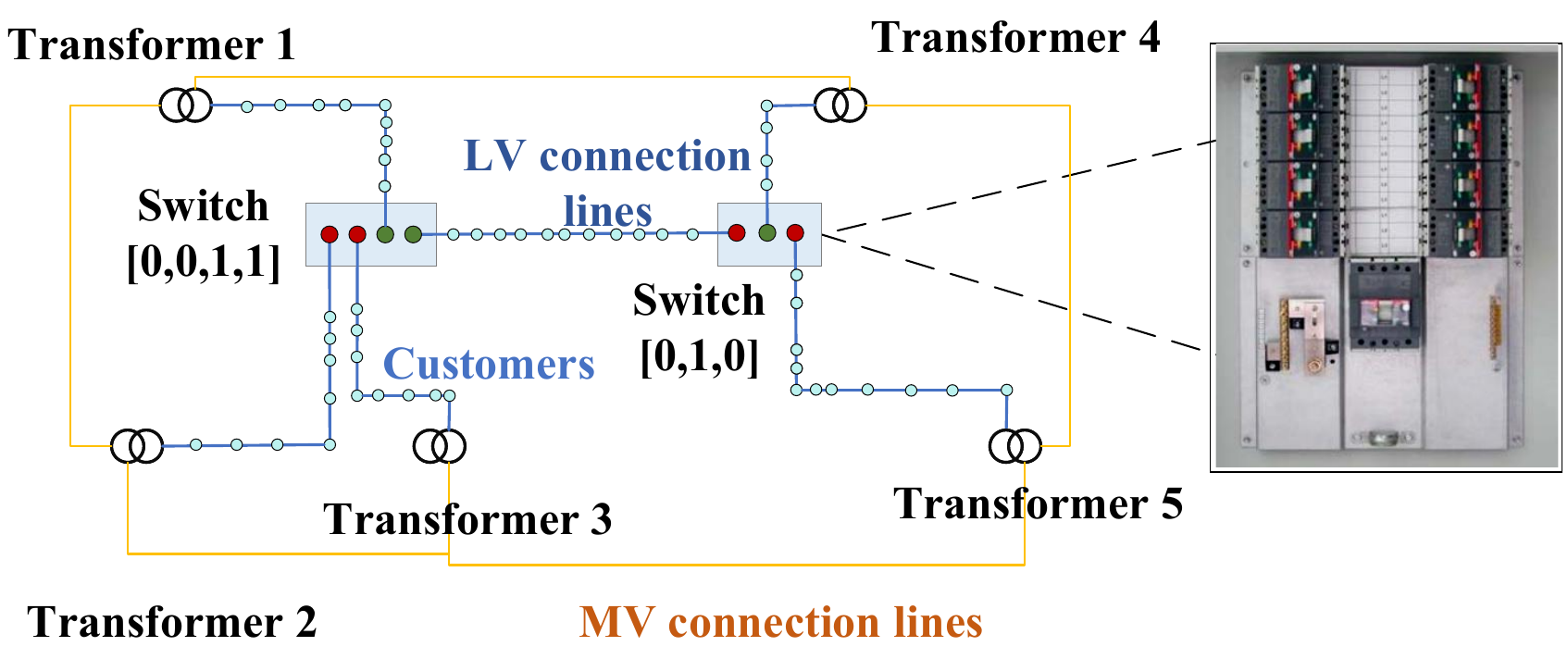}
\vspace{-0.1cm}
\caption{\textnormal{\textcolor{myblue}{Illustrative example of switch deployment and states in LVDNs: 1 represents the activated switch, and vice versa \cite{abb2025}.}}}
\vspace{-0.2cm}
\label{Fig.3}
\end{figure}

\vspace{-0.3cm}
\textcolor{myblue}{\subsection{Switch state identification}}
\label{2.1}
In LVDNs, multiple switch bars exist, each containing a different number of switches. Each switch connects two cables (e.g., A and B), which are linked to different transformers (e.g., AT and BT). Under normal conditions, the switch remains open. When transformer AT becomes overloaded, the switch is closed, shifting the load on cable A to transformer BT. Therefore, the switch state plays a crucial role in maintaining network efficiency and enabling effective load shifting. An example of switch deployment is illustrated in Fig.~\textcolor{myblue}{\ref{Fig.3}}. This non-uniform deployment poses challenges for accurately identifying the state of individual switches using SM data. To address this, a random forest (RF) algorithm is employed, leveraging voltage magnitude datasets to determine the state of switches within each switch bar. First, inspired by Label Encoding rules, the switch states in one bar are encoded into single-digital label formats, i.e., each combination of the switch states in the switch bar is uniquely mapped to a single digital label. The number of switches is assumed to be accessible for the DSO. For example, when there are 3 switches in the switch bar, the switch states $SS$ can be encoded as follows:
\begin{equation}
\label{eq1}
f(V) =
\begin{cases}
0, & \hspace{0.8cm} \text{if} \quad SS = [1, 1, 0] \\
1, & \hspace{0.8cm} \text{if} \quad SS = [1, 0, 1] \\
2, & \hspace{0.8cm} \text{if} \quad SS = [0, 1, 1]
\end{cases}
\end{equation}

After the switch states are encoded, the switch state identification problem in each bar is transformed into a standard multi-class classification task. The input for this classification task consists of time-series voltage magnitude datasets. However, the presence of missing data from SM or unmetered households results in varying numbers of available SMs on each feeder. Feeders with a higher number of available SM provide more data for the supervised learning process, potentially introducing bias in classification and leading to accuracy loss. Besides, when switch states remain unchanged for extended periods (e.g., a state $[1,0,1]$ resulting in an excessive number of samples labelled as $1$), the dataset becomes unbalanced. This imbalance further affects classification accuracy, posing a common challenge in machine learning.

To address the above-mentioned challenges, an SM selection process is employed before switch state identification. This selection process consists of two key steps:
\begin{itemize}
    \item \textit{Uniform SM number selection across feeders.}
    $M$ represents the minimum number of available SMs across all feeders. To ensure consistency, the number of SMs used for data extraction from each feeder is fixed at $M$. Specifically, the $M$ SMs closest to the switch bar are selected as representative features of that feeder. 
    \begin{equation}
    \label{eq2}
        M = \min \{m_1, m_2, \dots, m_K\},
    \end{equation}
    where $m_i$ is the number of available SMs on feeder $i$, and $K$ is the total number of feeders that are connected to the same switch bar.

    \item \textit{Balanced dataset construction.} Given the encoded switch state labels (i.e., 0, 1, 2, ...), an equal number of samples is selected for each category. For each encoded class $C_i$, the same number of samples $S$ is randomly selected.
    \vspace{-0.0cm}
    \begin{equation}
    \label{eq3}
        S = \min \{|C_1|, |C_2|, \dots, |C_J|\},
    \end{equation}
    where, $|C_j|$ represents the number of available samples for class $j$, and $J$ is the total number of classes.
\end{itemize}

Through this selection process, the impact of the varying number of available SMs and class imbalance is minimized, leading to a more balanced and effective training dataset for the multi-class classification task. The selected voltage measurements can be represented in a matrix form $V_{train}$, forming the input for the multi-class classification task. Each row in matrix $V_{train}$ represents training one sample with an encoded $SS$ label. $v^{s,C_j}_{m_k,M}$ represents the voltage magnitude of the house $M$ connected to the feeder with $m_k$ houses when the switch state belongs to category $C_j$.

\begin{equation}
V_{train} =
\begin{bmatrix}
\label{eq4}
v^{1,C_1}_{m_1,1} & \cdots & v^{1,C_1}_{m_1,M} & \cdots & v^{1,C_1}_{m_k,1} & \cdots & v^{1,C_1}_{m_k,M} \\
\vdots & & \vdots & & \vdots & & \vdots \\
v^{S,C_1}_{m_1,1} & \cdots & v^{S,C_1}_{m_1,M} & \cdots & v^{S,C_1}_{m_k,1} & \cdots & v^{S,C_1}_{m_k,M} \\
\vdots & & \vdots & & \vdots & & \vdots \\
v^{1,C_J}_{m_1,1} & \cdots & v^{1,C_J}_{m_1,M} & \cdots & v^{1,C_J}_{m_k,1} & \cdots & v^{1,C_J}_{m_k,M} \\
\vdots & & \vdots & & \vdots & & \vdots \\
v^{S,C_J}_{m_1,1} & \cdots & v^{S,C_J}_{m_1,M} & \cdots & v^{S,C_J}_{m_k,1} & \cdots & v^{S,C_J}_{m_k,M} \\
\end{bmatrix}
\end{equation}


\begin{figure}
\centering
\includegraphics[width=0.5\textwidth,height=0.21\textwidth]{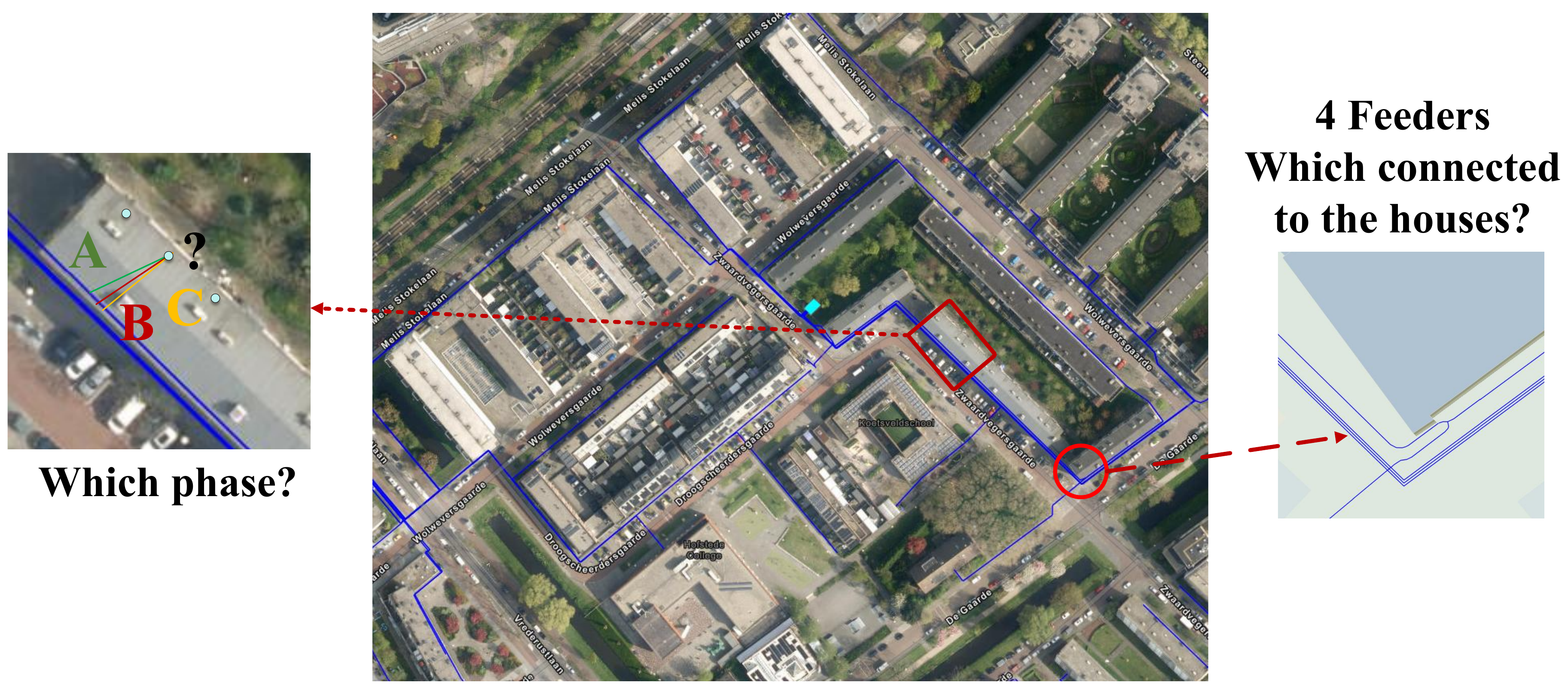}
\vspace{-0.3cm}
\caption{\textnormal{\textcolor{myblue}{A practical LVDN with multiple parallel feeders along the same street demonstrates the common challenge of ambiguous use-feeder connections and phase labelling, illustrating the physical complexity that makes accurate network topology identification (i.e., step (2) and (3)) difficult.}}}
\label{Fig4}
\end{figure}
The RF algorithm excels in multi-class classification by handling high-dimensional data, capturing complex feature interactions, and maintaining accuracy despite noise. Its ensemble learning approach—building diverse trees with random feature selection—prevents overfitting and enhances generalisation. RF also provides feature importance insights, making it ideal for switch state identification. Given a dataset $D = \{(X_1, y_1), \ldots, (X_n, y_n)\}$ where $X_i$ represents the feature vector (i.e., each row in  $V_{train}$) and $y_i$ is the encoded label by \textcolor{myblue}{(\ref{eq1})}, the general process of RF training read as:

\begin{enumerate}
    \item Draw \(B\) bootstrap samples from the training data $V_{train}$ and the corresponding labels.

    \item Grow an unpruned decision tree for each sample by randomly selecting a subset of features at each split.

    \item For classification, each tree votes on the most common class label, and the RF outputs the majority vote:
    \begin{equation}
    \label{eq5}
        \textcolor{myblue}{\hat{y} = \arg \max_j \sum_{b=1}^{B} \mathbf{I}(h_b(x) = j),}
    \end{equation}

\end{enumerate}
where $h_b(x)$ is the prediction of the $b$-th tree and \textcolor{myblue}{$\mathbf{I}(\cdot) $ is an indicator function that equals 1 if the condition is met and 0 otherwise.} Given a voltage sample $V_i$, each trained decision tree independently predicts $h_b(x)$ and the final labels of the switch are obtained by the majority voting mechanism.

\textcolor{myblue}{In the topology correction context, the use of supervised learning for switch state identification is practical. Unlike topology identification problems that assume a completely unknown network topology, topology correction relies on existing topology records and historical operation logs maintained by DSOs. Although these records may contain outdated information, they still provide useful prior knowledge about typical switch configurations. Historical voltage data collected under known operating conditions can therefore be associated with these configurations to form labelled samples for model training. Moreover, the number of feasible switch configurations within a feeder is usually limited. For example, in a switch cabinet with three switches, only 7 possible configurations exist, as shown in right part of Fig. \ref{Fig.3}. Therefore, supervised learning provides a feasible solution for identifying switch states within the proposed topology correction framework.}

\textcolor{myblue}{\subsection{User-feeder connections identification}}
\label{2.2}
The connection relations between customers and the feeders might be outdated and not always available for DSOs, which is ignored by multiple similar works~\cite{liu2025topology,10138375}, especially for houses and multiple-layer buildings that are close to the streets with multiple deployed underground feeders. A practical case is shown in Fig.~\textcolor{myblue}{\ref{Fig4}}. Those unclear user-feeder connections also impact the phase label identifications. A correlation approach is introduced to reveal the user-feeder connections, which only rely on time-series voltage magnitude datasets. Two practical cases are addressed: 1) user-feeder connection identification without reference and 2) user-feeder connection identification with limited, accurate recordings.

\textcolor{myblue}{\subsubsection{Connection identification without recordings}}
\label{sections2.2.1}
The proposed user-feeder connection identification algorithm combines the Hierarchical clustering (HC) algorithm and the correlation analysis based on the modified Pearson correlation coefficient (PCC). Algorithm \textcolor{myblue}{\ref{A1}} depicts the process of the proposed approach. The voltage magnitude dataset is stored in a matrix $V$, which is used as the input for the algorithm. Vector $V_n=[v_{n,1},v_{n,2},...,v_{n,\cal{T}}]^T$ represents the time series voltage magnitude at household $n$. $v_{n,t}$ represents the voltage magnitude at household $n$ at time $t$. $T$ is the measurement period of the SM data.
\begin{algorithm}[!t]
\caption{MFP-based User-Feeder Identification without Available Labels}
\label{A1}
\KwIn{$V$, $K_s$, $N$}
\For{$n \le N$}{
    \For{$m \le N$}{
        $\text{PCC}(V_n, V_m)$ via Eq. (\ref{eq7}) \\
        $ \text{FPCC}(\rho_{n,m})$ via Eq. (\ref{eq8}) \\
    }
}
\For{$i \gets 1$ \textbf{to} $N$}{
    \For{$j \gets i+1$ \textbf{to} $N$}{
        $D_{ij} \gets \text{MFP}(V_n, V_m)$ via Eq. (\ref{eq9})
    }
}
Obtain $\mathbf{D} \gets \{D_{ij}\}$ \\

   

Clusters $\mathcal{C}$: $\gets$ Traditional HC and $\mathbf{D}$\\
labels $\gets$ Assigned labels to households based on $\mathcal{C}$\\
\KwOut{labels, Cluster: $C_1$, ..., $C_{K_s}$}
\end{algorithm}

%
\vspace{-0.3cm}
\begin{equation}
    \label{eq7}
   PCC(V_{n}, V_{m}) = \frac{\sum_{t=1}^{T} (v_{n,t} - \bar{V_{n}})(v_{m,t} - \bar{V_{m}})}{\sqrt{\sum_{t=1}^{T} (v_{n,t} - \bar{V_{n}})^2} \sqrt{\sum_{t=1}^{T} (v_{m,t}- \bar{V_{m}})^2}} \quad 
\end{equation}
where $\bar{V_{n}}$ and $\bar{V_{m}}$ are the mean values of samples $V_{n}$ and $V_{m}$, respectively.

Usually, the PCC of voltage magnitudes in the same feeders shows higher values compared to those in different feeders~\cite{7098424}. To amplify the difference between voltage correlation from different feeders, a modified Fisher $z$-transformation function is employed:
\begin{align}
    \label{eq8}
    &FPCC(V_{n},V_{m}) =  \ln[\frac{1+PCC(V_{n},V_{m})}{1-PCC(V_{n},V_{m}) + \alpha}],
\end{align}
where $\alpha$ added to the denominator is used to avoid an infinite value of $FPCC(\cdot)$ and to control its distribution region.

To amplify the difference between voltage from different feeders, a modified $FPCC(\cdot)$ value (MFP) is employed to replace the Euclidean distance in the traditional HC algorithm.
\begin{align}    
    \label{eq9} 
    MFP(V_{n},V_{m}) = 1-min\{\frac{1}{4}\ln(1+e^{a+4\cdot FPCC(V_{n},V_{m})}),1\}
\end{align}

The second term in \textcolor{myblue}{(\ref{eq9})} is a modified likelihood function $F(\cdot)$~\textcolor{myblue}{\cite{garcia2023phase}}, which is used to calculate the $MFP(\cdot)$ between each sample $V_{n}$ and the rest of the samples. Since the range of the function $F(\cdot)$ is $(0, 1]$, the range of the distance metric $MFP(\cdot)$ is correspondingly $[0, 1)$. A value of $MFP(V_{n}, V_{m})$ closer to $0$ indicates a higher likelihood that the two samples $V_{n}$ and $V_{m}$ are collected from the same substation.

Given the switch state obtained from Section \ref{2.1}, the feeder-transformer connection is available, and the number of feeders connected to the same transformer is assumed to be accessible for the DSO and taken as the number of clusters (i.e., $K_s$). \textcolor{myblue}{Lines 1-7 in Algorithm  {\ref{A1}} are to obtain the MFP of the input voltage magnitude dataset. Lines 8-12 are to find the link matrix of matrix $D$. From line 13 onward, the input data are clustered into a $K_s$ cluster using the obtained linkage matrix.} If two users share the same label, they are connected to the same underground feeder, and vice versa.

\begin{algorithm}[!t]
\caption{MFP-based User-Feeder Identification with Available Labels}
\label{A 2}
\KwIn{$V$, $N_k$, $N_{un}$, $L_{un}$, $k$}
\For{ $n \geq N_{un}$}{
    \For{$m \le N_k$}{
        $D_0(n,m)$ $\gets$ $D$($V_n$,$V_m$) via Eq. (\ref{eq9})
    }
    $\hat{y} = \arg\max_{c} \sum_{n \in \mathcal{N}_k} \mathbb{I}(y_n = c)$ \\
    $L_{un}[n]$ $\gets$ $\hat{y}$\\
}
\KwOut{$L_{un}$ for households with unknown labels}
\end{algorithm}

\textcolor{myblue}{\subsubsection{Connection identification with recordings}}
When limited user-feeder connections are known, the user-feeder identification problem becomes more manageable compared to the fully unknown scenario. There are two sources of accurate recordings: 
\begin{itemize}
    \item Historical recordings from DSO's datasets.
    \item Identified connections with high confidence from Section~ \textcolor{myblue}{\ref{sections2.2.1}}.
\end{itemize}
\begin{figure*}
\centering
\includegraphics[width=1.0\textwidth,height=0.26\textwidth]{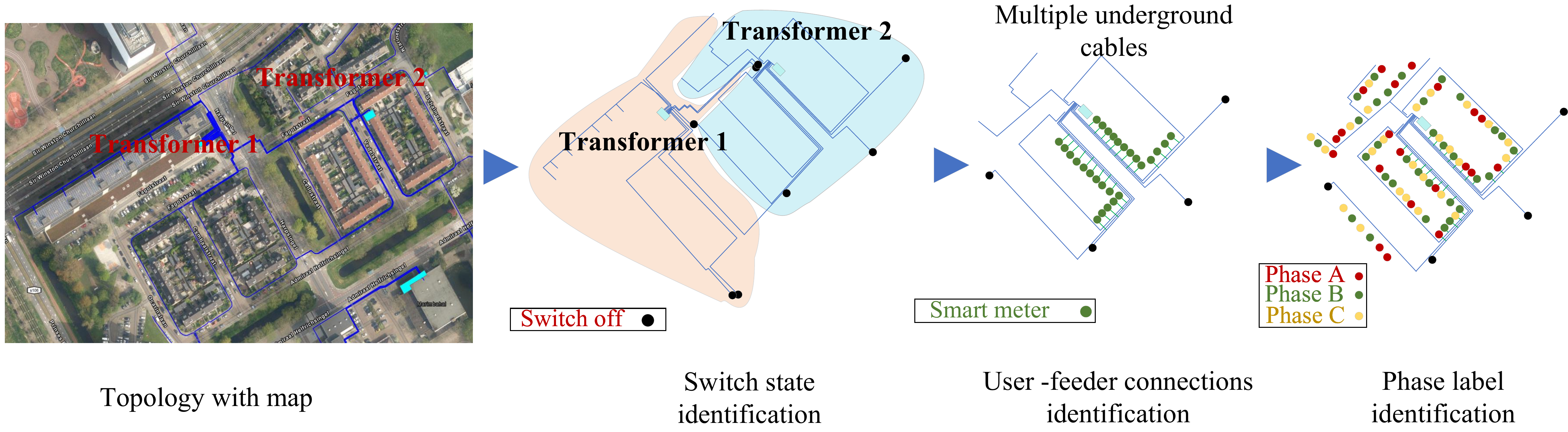}
\vspace{-0.2cm}
\caption{\textnormal{\textcolor{myblue}{Illustrative example showing the sequential process and intermediate outputs of the proposed topology correction framework, demonstrating (1) switch states, (2) user-feeder connections and (3) phase labels.}}}
\label{Fig.5}
\end{figure*}

Those available recordings transform the identification problem from an unsupervised learning problem to a supervised learning problem. Various supervised learning algorithms could be used to identify unknown connections, e.g., ANN, KNN, etc. The traditional KNN algorithm and the simplified MFP-based approach were employed and compared. The KNN algorithm is a widely used supervised learning method for classification tasks. It classifies an unknown sample by considering the labels of its nearest neighbours in the feature space, which is similar to identifying the user-feeder connection based on their location and the neighbours with known labels. Given a set of labelled training samples, the classification of a new sample is determined based on the majority class among its $ k $ nearest neighbours. The distance between samples is commonly measured using the Euclidean distance
\begin{equation}
    \label{eq10}
    d(V_n, V_m) = \sqrt{\sum_{t=1}^{T} (v_{n,t} - v_{m,t})^2}.
\end{equation}

Once the distances between the new sample and all labelled samples are computed, the $ k $ nearest neighbours are selected. The final classification is determined using majority voting.
\begin{equation}
\label{eq11}
    \textcolor{myblue}{\hat{y} = \arg\max_{c} \sum_{n \in \mathcal{N}_k} \mathbf{I}(y_n = c)}
\end{equation}

$\mathcal{N}_k$ denotes the set of $ k $ nearest neighbors, $y_i$ represents the label of neighbour $i$.
In user-feeder connection identification, the KNN algorithm assigns users without clear connections to the most probable feeder based on similarity to datasets from users who have available user-feeder connection recordings. Motivated by the traditional KNN approach, the proposed MFP-based approach is simplified to reveal the unknown user-feeder connection in DNs, which is depicted in Algorithm \textcolor{myblue}{\ref{A 2}}. $\mathcal{N}_{un}$ is the number of user with unclear connections, and $\mathcal{L}_{un}$ is their identified labels set.


\textcolor{myblue}{\subsection{Phase label identification}}



\begin{figure}
\centering
\vspace{-0.2cm}
\includegraphics[width=0.38\textwidth,height=0.55\textwidth]{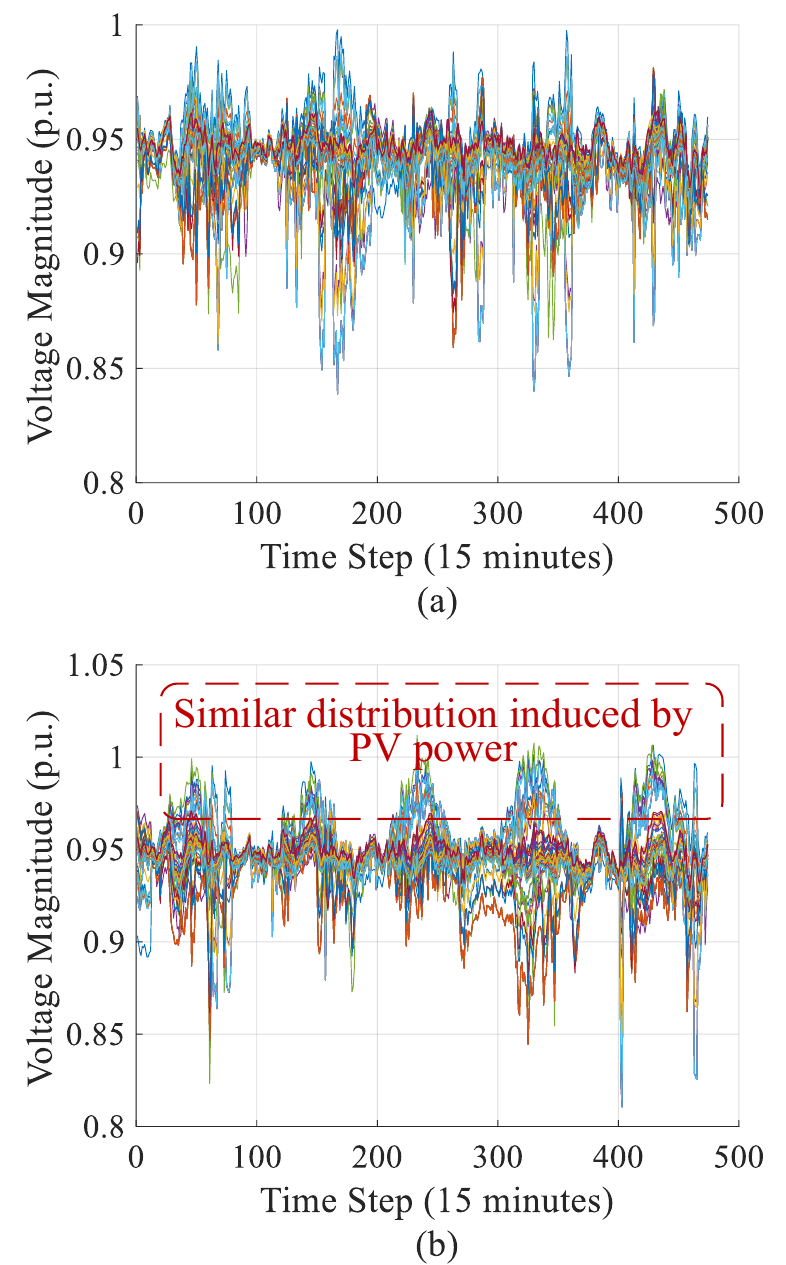}
\caption{\textnormal{\textcolor{myblue}{Distribution of voltage magnitudes in a LVDN:(a) winter datasets and (b) summer datasets.}}}
\vspace{-0.2cm}
\label{Fig10}
\end{figure}

After the second step in the proposed framework (i.e., user-feeder connection identification in Section \textcolor{myblue}{\ref{2.2}}), the third step is to identify the phase labels of the customer connection for each feeder, which is a typical 3-cluster classification problem. The phase identification approach is introduced by modifying Algorithm~\textcolor{myblue}{\ref{A1}}, i.e., removing Fisher $z$-transformation and using normalized voltage data $V^*$ as inputs.
\begin{equation}
V = 
\begin{bmatrix}
\label{eq6}
v_{1,1}& \cdots & v_{1,t} & \cdots & v_{1,T} \\
v_{2,1}& \cdots & v_{2,t} & \cdots & v_{2,T} \\
\vdots&   & \vdots  &  & \vdots  \\
v_{N,1}& \cdots & v_{N,t} & \cdots & v_{N,T}
\end{bmatrix} \hspace{1.0cm} 
\end{equation} 
\begin{equation}
\label{eq13}
    v_{n,t}^{*} = \frac{v_{n,t} - \mu_t}{\sigma_t}, \hspace{2.5cm} n\in \mathcal{N}, t \in \mathcal{T}
\end{equation}

$v_{n,t}$ is the value of the $t$-th feature for the $n$-th user. $\mu_t$ is the mean of the $t$-th feature, and $\sigma_t$ is the standard deviation of the $t$-th feature. Given the normalized voltage dataset $V^*$ and set $K_s$ as 3, modified Algorithm \textcolor{myblue}{\ref{A1}} is adopted to identify the phase labels of each house. The output of each step is depicted in Fig. \ref{Fig.5}, and the obtained topology presents the clear connections between users and transformers feeders, and phases.

\textcolor{myblue}{\subsection{Time-based data selection}}
\label{2.4}
Compared to traditional LVDNs, the voltage magnitude measurements in modern LVDNs may exhibit higher values due to the integration of PV systems. Within the same distribution network area, environmental factors such as light intensity, temperature, and humidity remain relatively uniform, leading to similar PV generation patterns among users. The voltage magnitudes from the same LVDN are depicted in Fig.~\ref{Fig10}. This effect is particularly pronounced during summer when high-power PV systems are connected (i.e., the voltage magnitude distribution in Fig.~\ref{Fig10}(b)). Consequently, users with identical PV capacities but connected to different phases may experience analogous voltage magnitude fluctuations. This phenomenon introduces challenges to the identification steps based on correlation analysis, as the voltage magnitude variations induced by PV generation may obscure the underlying phase-dependent characteristics.

Although various signal processing techniques, such as SA and high-pass filters, can be employed to mitigate high-frequency or low-frequency noise induced by household PV, the determination of appropriate parameters must be tailored to specific datasets. This requirement presents practical challenges in implementing these methods effectively. \textcolor{myblue}{Inspired by the power-band based strategies in~\cite{10533846}, a time-based data selection strategy is introduced to extract voltage measurements during periods when household loads dominate voltage variations, and PV generation remains lower than the local demand.} The proposed time-based data selection strategy is easily deployed and fast. Two strategies for selecting relevant data are introduced:

\begin{itemize}
    \item \textit{Fixed time threshold}. Data is selected from a predetermined time window, such as from 22:00 to 5:00, when PV generation is minimal.
    \begin{equation}
    \label{eq14}
    \mathcal{T}_d = \{t \mid t_1 \leq t \leq t_2\},
    \end{equation}
    where $t_1$ and $t_2$ define the fixed time window within which the data are extracted.
    \item \textit{Dynamic time threshold}. Data is selected based on historical PV output values, ensuring that the chosen data corresponds to periods when PV generation is lower than the user load.
    \begin{equation}
    \label{eq15}
    \mathcal{T}_d = \{t \mid P_{\text{PV}}(t) < P_{\text{load}}(t)\},
    \end{equation}
    where $\mathcal{T}_d$ represents the selected time instances, $P_{\text{PV}}(t)$ is the PV output at time $t$, and $P_{\text{load}}(t)$ is the user load at time $t$. Only the data points satisfying this condition are retained for phase identification.
\end{itemize}
\textcolor{myblue}{The proposed approach adaptively selects between Eq. \eqref{eq14} and Eq. \eqref{eq15} based on PV data availability. When PV forecasts are accessible, Eq. \eqref{eq15} is adopted to refine the selection by identifying intervals with low PV injection. Conversely, in the absence of such data, Eq. \eqref{eq14} serves as an easy and alternative by utilizing post-sunset voltage measurements, ensuring the model's applicability. By applying these time-based selection strategies, the impact of PV-induced voltage fluctuations can be mitigated, thereby improving the robustness of correlation-based phase identification methods.}

\textcolor{myblue}{For three-phase users, SMs are assumed to provide three independent voltage time-series corresponding to each phase. In the proposed methodology, each phase is modelled as an individual dataset and processed analogously to single-phase users. The algorithm performs independent analysis on the three phase-specific time-series and subsequently assigns an ordered set of phase labels, corresponding to one of the six feasible phase permutations (A–B–C, A–C–B, B–A–C, B–C–A, C–A–B, C–B–A). This utilisation way of three-phase datasets enables consistency with the single-phase data processing, avoiding inducing extra calculation to the approach.}

\section{\textcolor{myblue}{Case study}}
\label{case}

\textcolor{myblue}{To validate the proposed framework, this section conducts comprehensive evaluations across diverse Dutch grid scenarios. Unlike existing methods that often target isolated topology components, our approach provides an integrated solution for correcting switch states, user–feeder connections, and phase labels. Consequently, each step algorithm is benchmarked against representative machine learning techniques to demonstrate its feasibility and robustness. Furthermore, the following cases also evaluate the method's practical feasibility for DSO deployment under varying network scales and data distributions.}

The code used in this paper is available online~\footnote{https://github.com/distributionnetworksTUDelft/Data-driven-approach-for-topology-correction-in-LVDNs-}. The resolution of the dataset is 15 minutes, resulting in 96 data points per day. The base three-phase voltage level is 0.4 kV, and the loads of three-phase customers are non-uniformly distributed across Phases A, B, and C. The time thresholds $t_1$ and $t_2$ are set to 20 and 88. \textcolor{myblue}{To ensure data privacy, SM datasets provided by the DSO are anonymized before analysis. The original customer identifiers and location information are removed, and the SM data are randomly assigned to end-users within the studied networks. Then, the voltage magnitude profiles are generated using a power flow model, being solved by using the Power Grid Model package \cite{PowerGridModel}.}

\textcolor{myblue}{\subsection{Switch state analysis}}

\begin{figure}
\centering
\includegraphics[width=0.38\textwidth,height=0.58\textwidth]{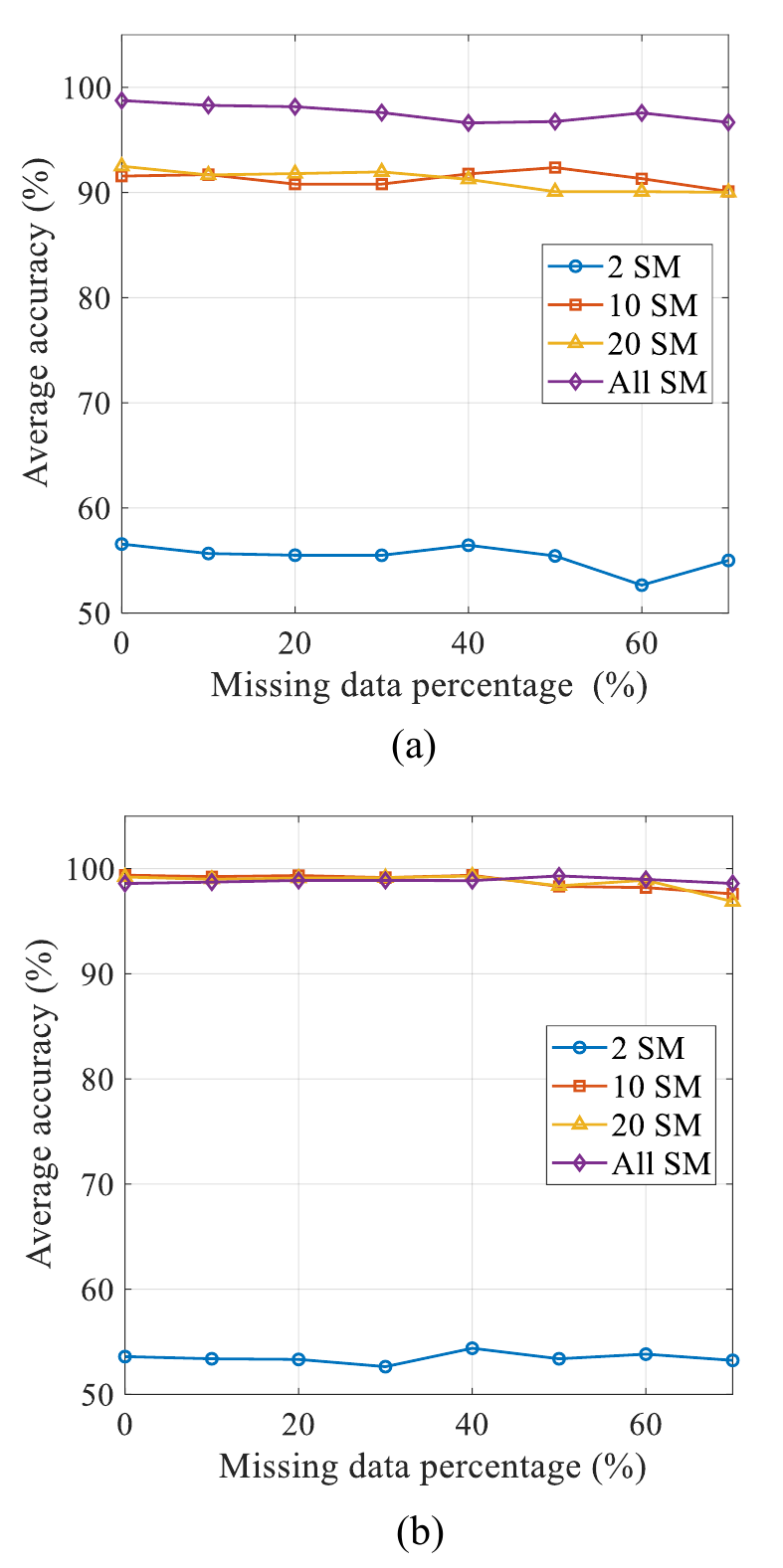}
\caption{\textnormal{\textcolor{myblue}{Switch state identification accuracy under incomplete dataset by: a) RF and b) SVM.}}}

\label{Fig6}
\end{figure}

\begin{table}
    \centering
    \renewcommand{\arraystretch}{1.2} 
    \setlength{\tabcolsep}{5pt} 
    \arrayrulewidth=0.8pt 
    \caption{Accuracy (\%) comparison under multiple scenarios.}
    \label{table2}
    \resizebox{0.41\textwidth}{!}{  

    \begin{tabular}{lcccc}
        \hline
        \hline
        \textbf{Number} & \multirow{1}{*}{\textbf{Number}} &  \multicolumn{3}{c}{\textbf{Methods}}  \\
        \cline{3-5}
        \textbf{of Switch} & \textbf{ of available SM} & \textbf{RF} & \textbf{SVM}  & \textbf{GNB} \\
        \hline
        \multirow{4}{*}{\shortstack{\textbf{3}\\\textbf{}}} & \multirow{1}{*}{2} & 69.16 & 63.31  & 60.39 \\
        & 10 & 87.98 &  88.63  &  85.71 \\
        & \multirow{1}{*}{20} & 88.31 &  88.96 &  84.74\\
        & All & 95.45 &  96.75  &  89.61  \\
        \hline
        \multirow{4}{*}{\shortstack{\textbf{5}\\\textbf{}}} & \multirow{1}{*}{2} & 56.56 &  53.59 & 56.72 \\
        & \multirow{1}{*}{10} & 91.56 & 99.38  &  95.16 \\
        & \multirow{1}{*}{20} & 92.50 & 99.21  &  94.06 \\
        & All & 98.75 &  98.59 & 95.94  \\
        \hline
        \hline
    \end{tabular}
    }
    \vspace{-0.2cm}
\end{table}

The switch bar is connected to three and five transformers, respectively, with all transformers assumed to be connected to the same MV network, which introduces similar low-frequency variations in customer voltage profiles. Each feeder connects to 49 customers. Using the voltage datasets selected through the time-based strategy described in Section \textcolor{myblue}{\ref{2.4}}, the performance of three models is evaluated under varying numbers of available SMs at each feeder, including RF, Support Vector Machine (SVM), and Gaussian Naive Bayes (GNB).

\begin{table*}[htbp]
\centering
\caption{\textcolor{myblue}{Quantitative description of test LVDN topologies.}}
\label{tab:topology_description}
\resizebox{0.75\textwidth}{!}{  
\begin{tabular}{c|c|c|c}
\hline
\hline
\textbf{Network} & \textbf{Feeder number} & \textbf{Nodes per feeder} & \textbf{Structure type} \\
\hline
CN1 & 2 & 20$\sim$26  & Complex + Complex \\
CN2 & 2 & 20$\sim$26  & Simple + Complex \\
SN  & 2 & 20$\sim$25  & Simple + Simple \\
SNL & 2 & 20$\sim$25  & Long Simple + Long Simple \\
SNB & 2 & 20$\sim$25  & Branched Simple + Branched Simple \\
\hline
Extended cases & 3 & 20--30  & Mixed (Simple/Complex) \\
\hline
\hline
\end{tabular}
}
\end{table*}

Table \textcolor{myblue}{\ref{table2}} demonstrates that the accuracy of all methods improves as the number of available SM increases. SVM achieves high accuracy (i.e., above 98\% for 5 switches scenarios and 88\% for 3 switches scenarios) once the number of houses reaches 10, while RF reaches the same level only when all SM data are available. This suggests that SVM is more efficient in learning from limited data, likely due to its ability to find optimal decision boundaries even with fewer samples. For the two SM data available scenarios, the accuracy of RF is about 5\% higher than that of SVM in two scenarios (i.e., 3 and 5 switches in the switch bar). Moreover, GNB underperforms compared to RF and SVM, with its accuracy lagging behind when the number of houses is low (lower than 90\% and 95\% under all the SMs data in two cases, respectively). 

Given datasets with random missed data points, the accuracy of RF and SVM were assessed, and the average accuracy under five simulations is shown in Fig. \textcolor{myblue}{\ref{Fig6}}. In general, the accuracy of the two approaches improves significantly as the number of available SMs increases and slightly decreases as the missing percentage increases, demonstrating the robustness of the proposed method with sufficient data availability. When only 2 SMs data are available, the accuracy remains consistently low (around 55\%–60\%), indicating the method’s limited capability under sparse data conditions. Besides, the sensitivity of SVM accuracy to the available data is less than that of RF, since the accuracy of SVM remains close when the number of available SM is larger than 10. These findings highlight the robustness of SVM and RF for switch state identification tasks in LVDNs.

\begin{figure}
\centering
\vspace{0.3cm}
\includegraphics[width=0.43\textwidth,height=0.48\textwidth]{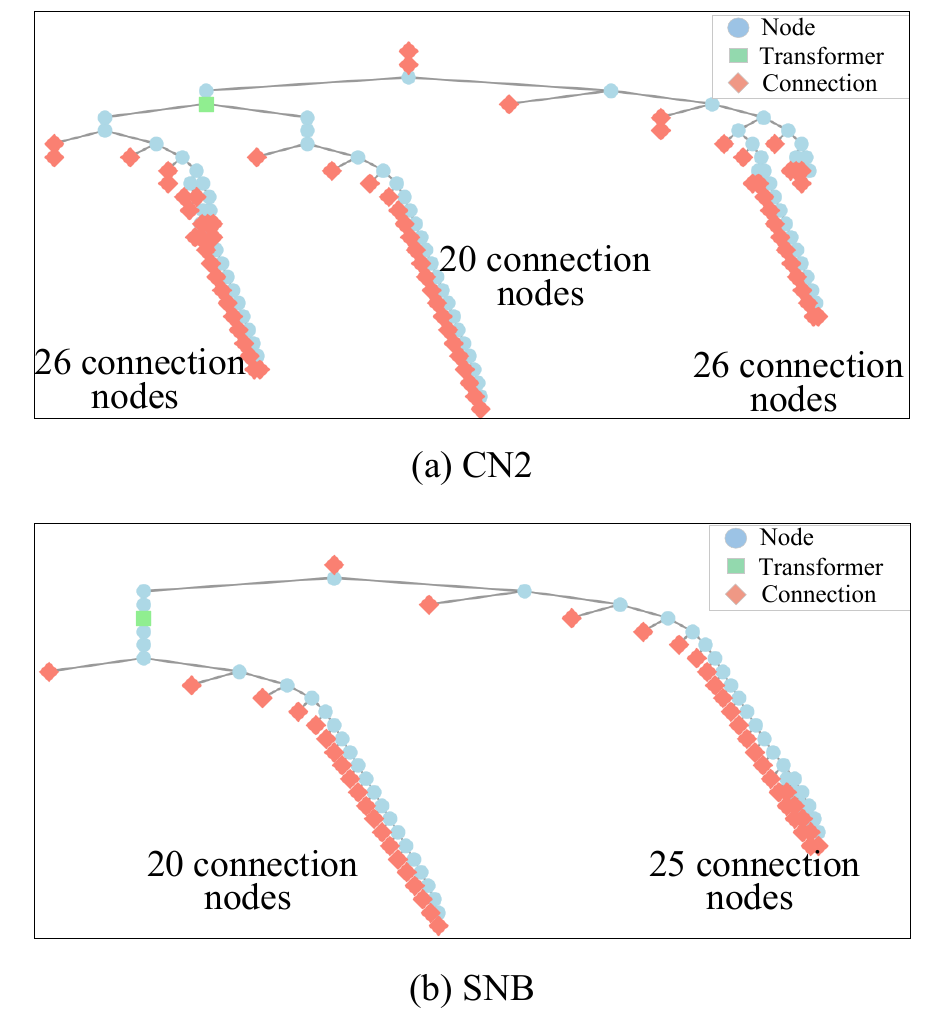}
\caption{\textnormal{\textcolor{myblue}{Topology of CN2 in (a) and SNB in (b).}}  }
\vspace{-0.2cm}
\label{Fig9}
\end{figure}
\begin{table}
    \centering
    \renewcommand{\arraystretch}{1.2} 
    \setlength{\tabcolsep}{5pt} 
    \arrayrulewidth=0.6pt 
    \caption{Accuracy (\%) comparison under multiple scenarios.}
    \label{table3}
    \resizebox{0.38\textwidth}{!}{  
    \begin{tabular}{clccccc}
        \hline
        \hline
        & \multirow{1}{*}{\textbf{Network}} & \multirow{2}{*}{\textbf{Phase}} & \textbf{Per-node} & \multicolumn{1}{c}{\textbf{Two}} & \multicolumn{1}{c}{\textbf{Three}} &\\
        & \textbf{Acronyms} & & \textbf{connection} & \textbf{feeders}  & \textbf{feeders}& \\
        \hline
        & \multirow{4}{*}{\shortstack{\textbf{CN1}\\\textbf{}}} & \multirow{2}{*}{1} & multiple & 100 & 100&\\
        \cline{4-7}
        & & & single & 100  & 96.42&\\
        \cline{3-7}
        & & \multirow{2}{*}{3} & multiple & 67.44 & 55.04& \\
        \cline{4-7}
        & & & single & 50.00  &  46.43 &\\
        \hline
        & \multirow{4}{*}{\shortstack{\textbf{CN2}\\\textbf{}}} & \multirow{2}{*}{1} & multiple & 100 & 98.26&\\
        \cline{4-7}
        & & & single & 100  & 100&\\
        \cline{3-7}
        & & \multirow{2}{*}{3} & multiple & 42.86  & 30.43&\\
        \cline{4-7}
        & & & single & 62.50  & 35.56&\\
        \hline
        & \multirow{4}{*}{\shortstack{\textbf{SN}\\\textbf{}}} & \multirow{2}{*}{1} & multiple &  94.56 & 67.56&\\
        \cline{4-7}
        & & & single & 75.56 & 38.46&\\
        \cline{3-7}
        & & \multirow{2}{*}{3} & multiple & 55.56  & 40.54&\\
        \cline{4-7}
        & & & single &  48.89 & 24.15&\\
        \hline
        & \multirow{4}{*}{\shortstack{\textbf{SNL}\\\textbf{}}} & \multirow{2}{*}{1} & multiple & 96.30 & 97.33&\\
        \cline{4-7}
        & & & single & 82.22 & 84.84&\\
        \cline{3-7}
        & & \multirow{2}{*}{3} & multiple & 59.26 & 37.33 &\\
        \cline{4-7}
        & & & single &  46.67 & 30.30&\\
        \hline
         & \multirow{4}{*}{\shortstack{\textbf{SNB}\\\textbf{ }}} & \multirow{2}{*}{1} & multiple & 95.45& 92.50 &\\
        \cline{4-7}
        & & & single & 80.00 & 88.57 &\\
        \cline{3-7}
        & & \multirow{2}{*}{3} & multiple & 55.56 & 43.18&\\
        \cline{4-7}
        & & & single &  48.49 & 45.71 &\\
        \hline
        \hline
    \end{tabular}
    }
\end{table}

\textcolor{myblue}{\subsection{User-feeder identification analysis}}
\label{3.2}
The influence of network complexity, phase configuration, and per-node connection on classification accuracy was first analysed. The per-node connection is the number of households connected to a point, e.g., a multi-unit building. \textcolor{myblue}{The test networks are used to represent LVDNs with varying structural characteristics in terms of feeder length and branching complexity. Five representative configurations are considered, as summarised in Table~\ref{tab:topology_description}: CN1, CN2, SN, SNL, and SNB.  The topology of CN2 and SNB is shown in~Fig.~\ref{Fig9}.}

\textcolor{myblue}{In this work, feeder structures are classified as simple or complex based on two key topological attributes: feeder length and the number of branches. Specifically, simple-structured feeders are characterised by relatively short radial paths with limited branching, while complex-structured feeders exhibit longer feeder lengths and multiple branching points, resulting in deeper topologies. In Fig.~\ref{Fig9}. The red diamond markers represent connection nodes, i.e., the electrical connection points between the main feeder and end-users. The SNB represents the SN with a higher number of lateral connections originating from the main cable, i.e., more branch points where users are connected.}

Given datasets with SM errors but without available recordings of user-feeder connections, Table~\textcolor{myblue}{\ref{table3}} presents the accuracy comparison under five modified configurations of LVDNs.


\begin{figure}
\centering
\includegraphics[width=0.5\textwidth]{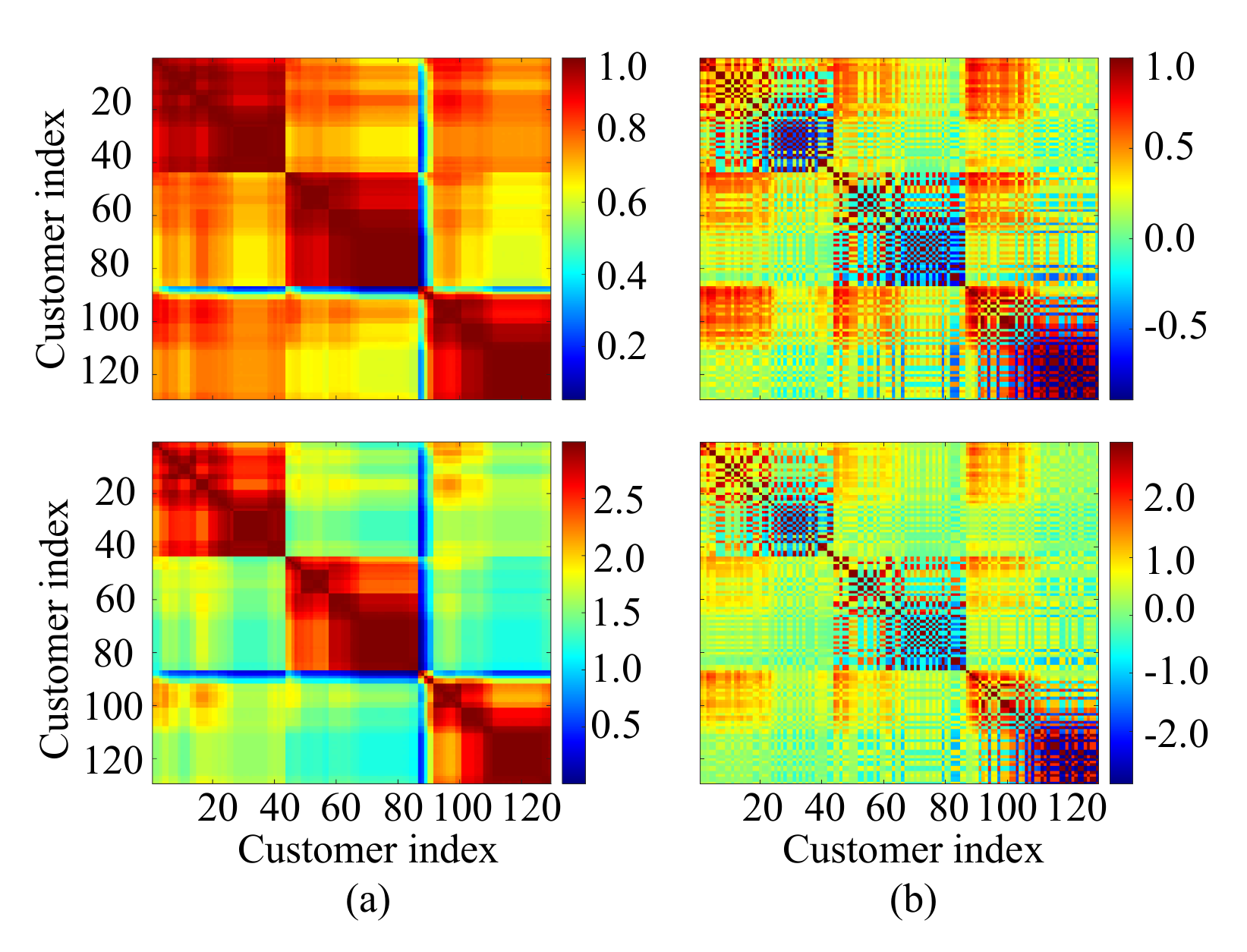}
\vspace{-0.4cm}
\caption{\textnormal{\textcolor{myblue}{PCC in first row and FPCC in second row among voltage datasets: a) with single-phase customers and b) with three-phase customers.}}} 
\vspace{-0.4cm}
\label{Fig7}
\end{figure}

\begin{table*}[h]
    \centering
    \renewcommand{\arraystretch}{1.2}
    \setlength{\tabcolsep}{5pt}
    \arrayrulewidth=0.6pt
    \caption{Average accuracy under three-phase datasets and multiple connections at nodes with SM errors.}
    \label{table4}
    \resizebox{0.75\textwidth}{!}{ 
    \begin{tabular}{lccccccccc}
        \hline
        \hline
        \multirow{1}{*}{\textbf{Network}} & \textbf{Unknown}  & \multicolumn{4}{c}{\textbf{Modified KNN}} & \multicolumn{4}{c}{\textbf{Traditional KNN}}\\
         \textbf{Acronyms} &  \textbf{rate} & $k=1$ & $k=3$ & $k=5$ & $k=8$ & $k=1$ & $k=3$ & $k=5$ & $k=8$\\
        \hline
        \multirow{4}{*}{\textbf{CN1}}
        & 10\% & 98.16\%&	98.10\%&	99.02\%& 99.59\%&  99.90\%	&99.79\%	&99.84\%	&99.95\%\\
        & 20\% &96.16\% &	96.49\% &	97.44\% &	98.94\% & 99.84\%&	99.67\%&	99.68\%	&99.51\% \\      
        & 30\% & 93.73\%&	95.19\%&	96.73\%&	98.65\% &99.52\%&	99.48\%&	99.08\%&	98.13\% \\
        & 40\% & 91.71\% &	92.79\% &	96.13\% &	97.84\% & 99.32\%&	98.75\%&	97.92\%	&94.70\%\\
        & 50\% & 88.84\% &	90.63\% &	95.59\% &	96.92\% & 99.03\%&	98.33\%	&94.76\%	&88.51\%\\
        \hline
         \multirow{4}{*}{\shortstack{\textbf{SN}\\\textbf{}}}  
        & 10\% & 97.63\%&	98.06\%&	98.41\%&	98.57\% &  99.65\%&	99.65\%&	99.20\%&	98.31\%\\
      
        & 20\% & 95.07\% &	95.80\% &	97.04\% &	97.11\% & 99.24\%&	99.30\%	&98.24\%	&96.98\%\\
        & 30\% & 92.19\% &	93.80\% &	95.20\% &	94.72\% & 98.98\%&	98.20\%	&97.00\%	&92.22\%\\
        & 40\% & 90.11\% &	91.37\% &	92.50\% &	92.44\% &98.56\%&	97.48\%&	93.54\%&88.94\%\\
        & 50\% & 87.11\% &	88.70\% &	89.43\% &	87.43\% &97.72\%	&92.98\%	&88.39\%	&83.50\%\\
        \hline
         \multirow{4}{*}{\shortstack{\textbf{SNB}\\\textbf{}}} 
        & 10\% & 97.43\%&	98.09\%&	98.33\%&	98.43\% & 99.61\%&	99.81\%	&99.48\%	&98.65\%\\
        & 20\% & 94.37\% &	95.87\% &	96.56\% &	96.83\% &99.39\%	&99.31\%	&98.63\%	&96.52\%\\
        & 30\% &91.80\% &	93.19\% &	94.78\% &	94.83\% & 98.81\% &	98.43\%	&96.69\%	&92.54\%\\
        & 40\% & 90.35\% &	91.70\% &	93.02\% &	91.87\% &98.43\%	&96.78\%&	93.57\%	&89.04\%\\
        & 50\% &87.50\% &	88.46\% &	89.31\% &	87.22\% &96.76\%	&94.15\%	&88.57\%	&82.11\%\\
        \hline
        \hline
    \end{tabular}}

\end{table*}

\textcolor{myblue}{\subsubsection{Identification without recordings}}
The PCC and MFP among customers in CN1 with multiple connections per node are depicted in Fig.~\textcolor{myblue}{\ref{Fig7}}.  As illustrated in Fig.~\textcolor{myblue}{\ref{Fig7}(a)}, the expression in~\textcolor{myblue}{(\ref{eq8})} reduces the voltage correlation between users connected to different feeders. This step helps distinguish between different connections more clearly. However, the complexity of the voltage data correlation shown in Fig.~\textcolor{myblue}{\ref{Fig7}(b)} makes it extremely challenging to simultaneously identify both the phase label and the user-feeder connection using only voltage data. A balanced LVDN represents a scenario where all users are connected to the same phase (i.e., a single-phase LVDN), whereas an unbalanced LVDN corresponds to a three-phase configuration.

Table \textcolor{myblue}{\ref{table3}} summarizes the accuracy of user-feeder connection identification across five LVDN scenarios. In general, the proposed approach demonstrates higher robustness to variations in the number of feeders connected to the same transformer and per-node connections but exhibits lower robustness in three-phase configurations. Specifically, accuracy drops from 100\% to 96.43\% in CN1 with three feeders and multiple connections per node, and from 94.56\% to 55.56\% in SN with two feeders and multiple connections per node, highlighting the impact of complex multi-phase settings. In simpler LVDNs, per-node connections significantly affect accuracy. For instance, in SNL with two feeders in a single-phase setting, accuracy declines from 96.3\% to 82.22\%, while in SNL with two feeders in a three-phase setting, it drops from 59.26\% to 46.67\%. In contrast, for complex LVDNs, accuracy remains above 95\% regardless of single or multiple per-node connections, indicating greater resilience in more intricate network structures. Moreover, accuracy in complex LVDNs is slightly higher than in simpler LVDNs, primarily due to feeder length differences. Simple LVDNs, which are modified from urban areas with high population density, have houses located close to each other (i.e., smaller distances between users with different connections). This proximity reduces the difference in the voltage variations between neighbouring houses, leading to greater identification accuracy. However, the length of the feeder and the number of branches have a minor effect on the accuracy, which is verified in the three simple LVDNs (i.e., SN, SNL, and SNB). 

\textcolor{myblue}{It is observed that the identification accuracy of the proposed method declines when applied to three-phase networks featuring three or more outgoing cables. This performance degradation highlights an inherent limitation of relying solely on voltage profiles in scenarios with high branch density and complex coupling. To overcome this shortcoming, incorporating multi-source datasets represents a promising avenue for future enhancement. Leveraging such auxiliary information can expand the feature space, thereby significantly improving the discriminative power and spatial resolution of the algorithm in more sophisticated grid structures.}

Since unbalanced LVDNs are more representative of real-world conditions compared to balanced LVDNs, only half of the identified user-feeder connections with high confidence are selected for further processing in the subsequent section. These experimental results highlight the inherent challenge of simultaneously identifying user-feeder connections and accurately determining the phase labels of individual users. The complexity arises from the intertwined nature of voltage variations across different feeders and phases, further complicated by factors such as DERs and seasonal load fluctuations. 

\textcolor{myblue}{\subsubsection{Identification with recordings}}
Given datasets with varying levels of unknown user-feeder connections, Table \ref{table4} compares the average accuracy of 20 simulations at each scenario (i.e., each unknown level). The results provide insights into the robustness and effectiveness of both methods across different network configurations. In general, as the percentage of unknown user-feeder connections increases from 10\% to 50\%, a gradual decline in accuracy is observed across all methods and network configurations. However, both the proposed MFP-based method and KNN (with varying $k$ values) maintain relatively high performance, with accuracy exceeding 91\% even at 30\% unknown connections.


\textcolor{myblue}{Table~\ref{table4} shows that the traditional Euclidean distance-based KNN consistently outperforms the proposed modified KNN when the neighborhood size is small (e.g., $k<8$), particularly under low levels of missing data. This behavior is expected, as a smaller $k$ relies on highly local information, where the standard distance metric can effectively capture similarity when sufficient accurate observations are available. However, this advantage diminishes as the proportion of missing data increases and the neighborhood size becomes larger. Under such conditions, the modified KNN demonstrates improved accuracy, particularly at higher unknown rates (e.g., 50\%) and larger $k$, where it achieves accuracy levels exceeding 87\%. This indicates that the proposed distance formulation is more effective in mitigating the distortion introduced by incomplete measurements, thereby preserving identification performance when reliable neighbor information becomes sparse. It is important to note that the selection of $k$ is inherently data-dependent in practical DSO applications. In data-sparse scenarios—where only a limited number of neighboring users have reliable phase labels—a small $k$ (e.g., $k=1$–$3$) is typically preferred, in which case the traditional KNN may be sufficient. In contrast, in more realistic settings with larger available datasets, DSOs often rely on a larger neighborhood to improve statistical reliability. In such cases, the proposed modified KNN provides clear advantages in terms of robustness and stability. This trend is further confirmed in the SNB network, where increased branching introduces additional structural complexity. While the traditional KNN shows a slight advantage at intermediate unknown rates (e.g., 15\%–20\%) and small $k$, the modified KNN achieves comparable or superior performance as missing data increases or when larger neighborhoods are considered. Overall, these results suggest that the proposed method is not intended to replace traditional KNN in all scenarios, but rather to enhance accuracy and options under realistic conditions characterized by uncertain neighborhood information. This flexibility allows DSOs to select or adapt the method based on data availability and operational requirements.}

\begin{figure}
\centering
\includegraphics[width=0.46\textwidth]{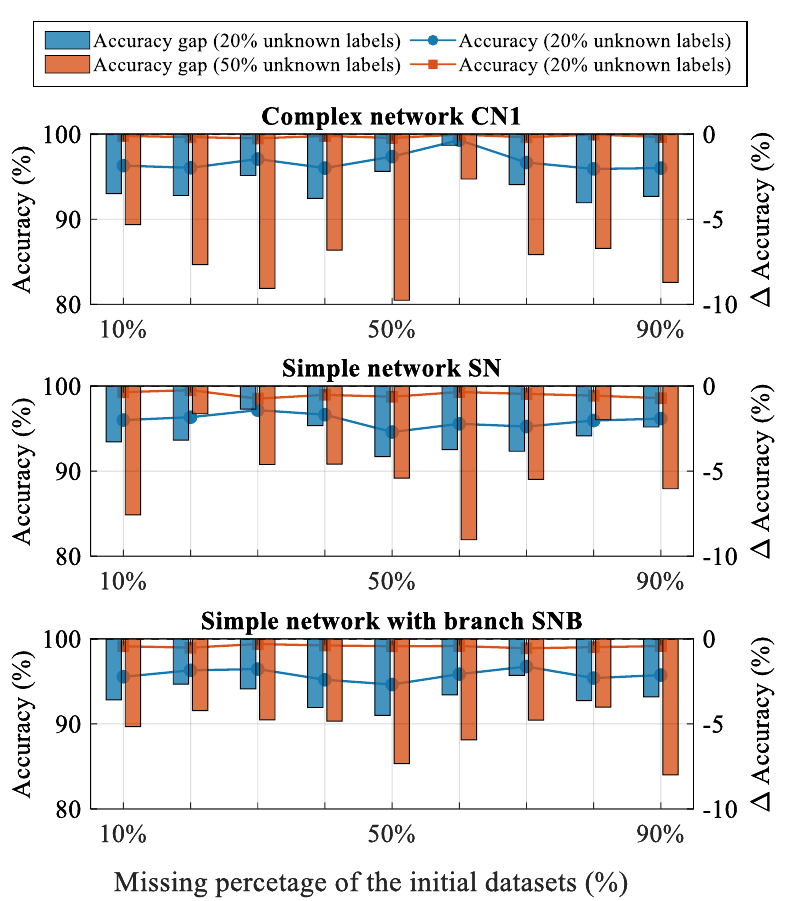}
\caption{\textnormal{\textcolor{myblue}{Mean accuracy and corresponding performance gap under varying levels of data incompleteness ($k = 3$).}}} 

\label{Fig8}
\end{figure}

\textcolor{myblue}{To validate the robustness of the proposed method with respect to network topology, missing data, and measurement noise, a comprehensive set of experiments was conducted under varying scenarios. Specifically, performance was evaluated across different levels of incompleteness of the data, varying proportions of available labeled data, and different noise magnitudes. The missing data ratio varied from $10\%$ to $90\%$. Regarding SM noise, four classes of SMs were considered according to IEC 62053-21, corresponding to error levels of $0.2\%$, $0.5\%$, $1\%$, and $2\%$ (extended to $5\%$ for extreme scenarios). Besides, to compare the accuracy between the traditional KNN and the improved KNN under different missing data conditions, an \emph{accuracy gap} ($\Delta$ accuracy) was introduced, defined as the difference between the average accuracy of the modified KNN algorithm and that of the traditional KNN algorithm. In this analysis, the number of neighbours was fixed at $k=3$. The corresponding results are illustrated in Fig. \textcolor{myblue}{\ref{Fig8}}.}

\textcolor{myblue}{From the line plots in Fig. \textcolor{myblue}{\ref{Fig8}}, it can be observed that the accuracy of the proposed method remains relatively stable across different levels of missing data, indicating strong robustness against data incompleteness. Furthermore, the bar charts show that the accuracy gap is generally negative under the considered noise levels when $k=3$, suggesting that the traditional KNN slightly outperforms the improved method in these scenarios. Nevertheless, the performance difference remains within $10\%$, which is consistent with the observations reported in Table \textcolor{myblue}{\ref{table3}}.}

\begin{figure}
\centering
\includegraphics[width=0.48\textwidth]{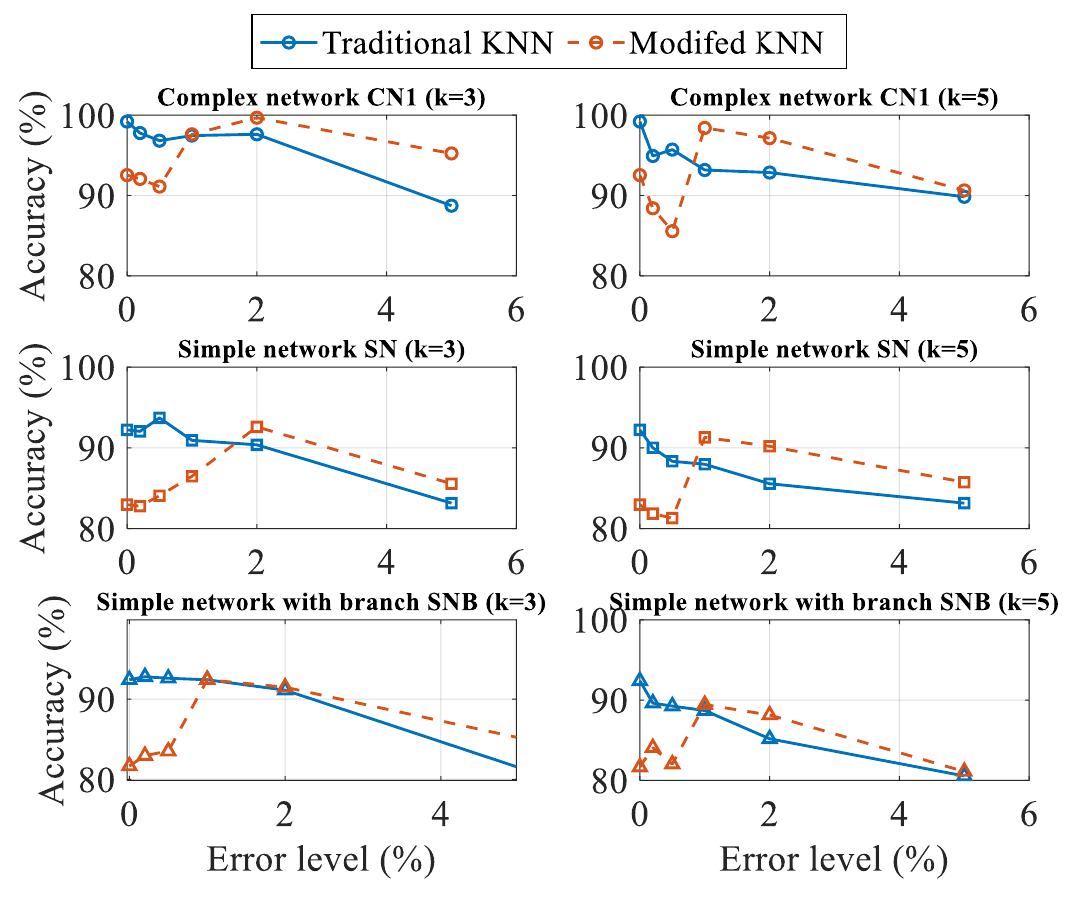}
\vspace{-0.4cm}
\caption{\textnormal{\textcolor{myblue}{Mean accuracy under multiple datasets with different noise level.}}} 
\vspace{-0.1cm}
\label{Fig11}
\end{figure}

\textcolor{myblue}{Fig. \textcolor{myblue}{\ref{Fig11}} presents the comparison under different noise conditions. As the noise level increases, the accuracy of the traditional KNN gradually decreases, indicating its sensitivity to measurement noise due to its reliance on Euclidean distance. In contrast, the proposed method exhibits a different trend. The accuracy initially slightly increases and then decreases slightly as noise grows, while maintaining comparable overall performance. This is because the proposed method is based on correlation analysis rather than Euclidean distance, making it inherently more robust to noise. Moderate magnitude noise can enhance the distinguishability between different users due to heterogeneous noise, leading to a temporary improvement in performance. However, as the noise magnitude becomes large, the similarity induced by high-amplitude Gaussian noise impacts correlation coefficient calculation, leading to accuracy reduction.}

\textcolor{myblue}{Based on Table \textcolor{myblue}{\ref{table3}} and Fig. \textcolor{myblue}{\ref{Fig8}}--\ref{Fig11}, it can be concluded that the two methods exhibit complementary characteristics under different conditions. For scenarios involving high noise levels or larger $k$ values, the improved KNN tends to achieve better performance. Conversely, under low-noise conditions and smaller $k$, the traditional KNN can serve as a competitive alternative. Therefore, an adaptive selection between the two methods can be employed to ensure reliable accuracy across diverse operating conditions.}

\textcolor{myblue}{\subsection{Phase identification analysis}}
The influence of feeder complexity, environmental conditions, and load distribution on identification accuracy is analyzed, with results summarized in Table \textcolor{myblue}{\ref{table5}} and \textcolor{myblue}{\ref{table6}}. CC\_PV: Represents a complex-structured feeder from CN1, with photovoltaics (PV) installed at each house. SC\_PV: Corresponds to a simple-structured feeder from SN, also with PV at each house. CC and SC: Represent feeders from CN1 and SN with PV, but with datasets preprocessed using the method described in Section \textcolor{myblue}{\ref{2.4}}. These four feeders serve single-phase users only.CC\_3\_NU and SC\_3\_NU: Denote complex and simple feeders equipped with three-phase meters, where the load is non-uniformly distributed among the three phases. “+20” indicates an additional 20 customers connected to the feeder. “Long feeder” refers to an extended feeder length compared to the original configuration. Accuracy and clustering purity are used to evaluate the performance of phase identification, as formulated in the expression \textcolor{myblue}{(\ref{eq16})}.
\begin{equation}
\label{eq16} 
P_{pu} = \frac{1}{N} \sum_{\mathcal{L}_k\in {\mathcal{L}}}\underset{i}{\operatorname{max}}|\mathcal{L}_k\cap \mathcal{L}_i| \ 
\end{equation} 

where $\mathcal{L}$ denotes the set of obtained labels for all input data, and $\mathcal{L}_k$ represents the label assigned to the data in cluster $\mathcal{C}_k$. $\mathcal{L}_i$ is the label set of data points that truly belong to class $\mathcal{C}_i$.

\begin{table}[!t]
\caption{\textnormal{Average purity (\%) under datasets form feeders with different components.}}
\label{table5}
\centering
\resizebox{\columnwidth}{!}{
\begin{tabular}{cccccccc}
\hline
\hline
\multirow{2}{*}{\textbf{Network}} & \multirow{2}{*}{\textbf{Setting}} &
\multicolumn{2}{c}{\textbf{Winter Dataset}} & \multicolumn{2}{c}{\textbf{Summer Dataset}} &\\
 &  & single & multiple & single & multiple \\
\hline
\multirow{2}{*}{CC\_PV}
& - & 100 & 100& 78.57 & 65.12 \\
& +20 & 100 & 100 & 77.08 &77.78   \\
\hline
\multirow{2}{*}{SC\_PV}
& -& 96.00& 97.06& 84.00 & 88.24\\
&longer feeder &  100 & 94.12 & 84.00& 88.24 \\
\hline
\multirow{2}{*}{CC}
& -& 100 &  100 & 100 & 97.67 &  \\
& +20& 100 &  100 & 100& 100 & \\
\hline
\multirow{2}{*}{SC}
&- & 88.88 &88.24& 68 &82.35\\
& longer feeder  & 100&100 & 68.00   &82.35\\
 \hline
\multirow{2}{*}{CC\_3\_NU}
& PV & 100 & 100 & 79.41 & 83.67& \\
&filtered & 100 & 100 & 100 & 100& \\
\hline
\multirow{2}{*}{SC\_3\_NU}
& PV &  97.14 & 97.92 & 71.43 & 100\\
& filtered &  97.92 & 94.29 &  100 & 95.83\\
\hline
\hline
\end{tabular}
}
\end{table}
 
Overall, Table \ref{table5} indicates consistently higher accuracy in winter datasets compared to summer datasets. This discrepancy is primarily due to voltage variations caused by household photovoltaic (PV) systems, which negatively impact phase identification performance. In the first four cases, where only single-phase or balanced three-phase customers are considered, the influence of PV-induced voltage fluctuations is evident. For instance, in the summer dataset, the CC\_PV feeder (a complex feeder from CN1 without PV filtering) achieves 88.89\% accuracy with a single connection per node but drops to 72.08\% with multiple connections per node. However, when PV-induced variations are filtered out—as in the CC and SC cases—accuracy consistently exceeds 95\%, highlighting the critical role of PV filtering in improving phase identification.

These findings demonstrate that phase identification accuracy is highly dependent on seasonal variations and dataset complexity, particularly in the presence of PV-induced voltage fluctuations. Without applying the proposed time-based selection process to mitigate PV impacts, accuracy remains below 90\% in most cases. In contrast, once PV-induced variations are filtered, accuracy consistently exceeds 95\%, regardless of network complexity or three-phase load distribution. Additionally, factors such as the number of connections per node, feeder length, and the number of connected houses have a relatively minor impact on classification accuracy compared to PV output power fluctuations. This underscores the importance of addressing PV-induced voltage variations to achieve robust and reliable phase identification.

\textcolor{myblue}{\subsection{Limitation analysis}}

\textcolor{myblue}{Based on the above discussion and simulation result analysis, it is verified that the proposed method offers an effective solution for DSOs to maintain accurate topology within Digital Twin frameworks via non-intrusive voltage datasets. However, its broad application faces three potential challenges:}
\begin{itemize}
    \item \textcolor{myblue}{Network Topology Complexity: The identification performance is increasingly challenged by sophisticated distribution structures, particularly in networks featuring more than three cables (e.g., high branch density), which complicates the correlation patterns.}
    \item \textcolor{myblue}{Data quality heterogeneity: according to the robustness analysis in Section \ref{3.2}, inconsistent SMs precision poses difficulties in tuning the hyperparameter $k$.}
    \item \textcolor{myblue}{Data granularity: if users only share the maximum/minimum voltage value \cite{liu2025topology}, the application of the developed approach will be significantly limited.}
    \item \textcolor{myblue}{Evolving load dynamics: The proliferation of heat pumps may change the distribution of winter data, potentially introducing new patterns of similarity that challenge the feasibility of the proposed approach.}

\end{itemize}

\textcolor{myblue}{In future work, these limitations will be further addressed. Meanwhile, the ongoing transition toward advanced smart meters (SMs) with enhanced sensing and communication capabilities provides significant opportunities to further improve the accuracy and practical applicability of the proposed approach in increasingly digitised power grids.}

\begin{table}[!t]
\caption{\textnormal{Average accuracy (\%) under datasets form feeders with different components.}}
\label{table6}
\centering
\resizebox{\columnwidth}{!}{
\begin{tabular}{cccccccccc}
\hline
\hline
\multirow{2}{*}{\textbf{feeder}} & \multirow{2}{*}{\textbf{setting}} &
\multicolumn{2}{c}{\textbf{Winter Dataset}} & \multicolumn{2}{c}{\textbf{Summer Dataset}} &\\
 &  & single & multiple & single & multiple \\
\hline
\multirow{2}{*}{CC\_PV}
&- & 100 & 100& 88.89 & 72.08\\
&+20 & 100 & 100&  76.11 & 87.88\\
\hline
\multirow{2}{*}{SC\_PV}
&- & 96.97& 97.78& 87.88 & 90.30\\
&longer feeder &  100& 94.75 & 87.88 & 90.30\\
\hline
\multirow{2}{*}{CC}
&- &100  &  100 & 100 & 97.78 & \\
& +20& 100 &  100 & 100 &100  & \\
\hline
\multirow{2}{*}{SC}
& -& 87.45& 87.88& 71.86 &79.22\\
& longer feeder & 100& 100& 71.86 &79.22\\
 \hline
\multirow{2}{*}{CC\_3\_NU}
& PV & 100& 100 & 86.11 & 85.19\\
&filtered &  100& 100 & 100& 100\\
\hline
\multirow{2}{*}{SC\_3\_NU}
& PV & 97.78  & 97.62 & 76.97 & 100\\
& filtered & 97.62 & 95.56 & 100 & 95.24\\
\hline
\hline
\end{tabular}
}
\end{table}

\section{Conclusion}
\label{Conclusion}
In this paper, we proposed a practical topology correction framework for LVDNs, incorporating a time-based smart meter selection strategy to mitigate the impact of PV systems and a balanced dataset construction to improve training effectiveness. The experimental results demonstrate that the proposed framework is highly robust, maintaining high accuracy even with incomplete datasets, regardless of random missing data, data dimensionality, or topological complexity. Specifically, the correlation-based user and phase identification shows strong resilience to data loss. However, the switch state identification component, based on supervised learning, is more sensitive to missing and asynchronous data. Moreover, the framework's performance is influenced by seasonal variations and voltage fluctuations caused by PVs, highlighting the importance of time-based data selection in ensuring reliable topology identification. Future extensions of this work will account for the influence of various flexible assets to enhance the scalability and precision of the framework.

\printcredits
\bibliographystyle{elsarticle-num}

\bibliography{ref}

\end{document}